\def \ntces {20,367}        
\def \nntltces {13,283}  
\def \nfpkoitces {2,786}  
\def \npctces {4,298}  
\def \npckois {4,293}  
\def \npcsystems {3,355}  
\def \npcmultisystems {636}  
\def \npcmultikois {1,574}  
\def \nkois {7,470} 
\def \nsingkois {5,864}  
\def \npcsingkois {2,661}  
\def \nfpsingkois {3,023}  
\def \nmultikois {1,786}  
\def \npcmultikois {1,632}  
\def \nfpmultikois {154}  
\def \nephemmatch {1,910}   
\def \nonlyephemmatch {189} 
\def \nfedkois {5,992}      
\def \ntlfedkois {5,854}     
\def \ntlfedkoisrvtl {5,700}  
\def \npredrtwentyfourkois {7,348}  
\def \npredrtwentyfourtlkois {6,491}  
\def \npredrtwentyfourfedpcs {3,772}   
\def \npredrtwentyfourfedpcsrvpc {3,654}   
\def \npredrtwentyfourfedfps {2,220}   
\def \npredrtwentyfourfedfpsrvfp {1,983}   
\def \nnewkois {1,478}      
\def \ntotkois {8,826}      
\def \nnewpcs {402}         
\def \npctofp {118}         
\def \nfptopc {237}         
\def \ntotpcs {4,696}       
\def \nkebs {2,605}         
\def \ninjecttces {42,264}  
\def \nnooffset {35,917}    
\def \nconfirmed {985}  
\def \nconfirmedpcs {976}  
\def \nebexclude {1,033}  
\def \nbankois {25}  

\def \rsun {$R_{\sun}$}

\def \rstar {$R_{\star}$}

\def \tstar {$T_{\star}$}
\def \rp {$R_{p}$}

\def \sp {$S_{p}$}
\def \teq {T$_{\rm eq}$}

\def \re {$R_{\earth}$}

\def \se {$S_{\earth}$}
\def \kepler {\emph{Kepler}}
\def \KEPLER {\emph{KEPLER}}
\def \keplers {\emph{Kepler's}}
\def \th {$^{\rm th}$}

\documentclass[apj,twocolappendix,numberedappendix]{emulateapj}

\let\underscore\_
\renewcommand{\_}{\discretionary{\underscore}{}{\underscore}}

\begin{document}

\title{Planetary Candidates Observed by \KEPLER{}. VII. The First Fully Uniform Catalog Based on The Entire 48 Month Dataset (Q1--Q17~DR24)}
\author{Jeffrey L. Coughlin\altaffilmark{1,2,3}, F. Mullally\altaffilmark{2,3}, Susan E. Thompson\altaffilmark{2,3}, Jason F. Rowe\altaffilmark{2,3}, Christopher J. Burke\altaffilmark{2,3}, David W. Latham\altaffilmark{4}, Natalie M. Batalha\altaffilmark{3}, Aviv Ofir\altaffilmark{5}, Billy L. Quarles\altaffilmark{3}, Christopher E. Henze\altaffilmark{3}, Angie Wolfgang\altaffilmark{6,15}, Douglas A. Caldwell\altaffilmark{2,3}, Stephen T. Bryson\altaffilmark{3}, Avi Shporer\altaffilmark{7}, Joseph Catanzarite\altaffilmark{2,3}, Rachel Akeson\altaffilmark{8}, Thomas Barclay\altaffilmark{3,9}, William J. Borucki\altaffilmark{3}, Tabetha S. Boyajian\altaffilmark{10}, Jennifer R. Campbell\altaffilmark{2,11}, Jessie L. Christiansen\altaffilmark{8}, Forrest R. Girouard\altaffilmark{3,12}, Michael R. Haas\altaffilmark{3}, Steve B. Howell\altaffilmark{3}, Daniel Huber\altaffilmark{2,13,14}, Jon M. Jenkins\altaffilmark{3}, Jie Li\altaffilmark{2,3}, Anima Patil-Sabale\altaffilmark{3,11}, Elisa V. Quintana\altaffilmark{2,3}, Solange Ramirez\altaffilmark{8}, Shawn Seader\altaffilmark{2,3}, Jeffrey C. Smith\altaffilmark{2,3}, Peter Tenenbaum\altaffilmark{2,3}, Joseph D. Twicken\altaffilmark{2,3}, Khadeejah A. Zamudio\altaffilmark{3,11}
}

\journalinfo{Accepted for Publication in the Astrophysical Journal on February 15, 2015}
\submitted{}

\altaffiltext{1}{\scriptsize jeffrey.l.coughlin@nasa.gov; jcoughlin@seti.org}
\altaffiltext{2}{\scriptsize SETI Institute, 189 Bernardo Ave, Suite 200, Mountain View, CA 94043, USA}
\altaffiltext{3}{\scriptsize NASA Ames Research Center, M/S 244-30, Moffett Field, CA 94035, USA}
\altaffiltext{4}{\scriptsize Harvard-Smithsonian Center for Astrophysics, 60 Garden Street, Cambridge, MA 02138, USA}
\altaffiltext{5}{\scriptsize Department of Earth and Planetary Sciences, Weizmann Institute of Science, 234 Herzl St., Rehovot 76100, Israel}
\altaffiltext{6}{\scriptsize UC Santa Cruz, Department of Astronomy, 1156 High Street, MS: UCO / LICK, Santa Cruz, CA 95064, USA}
\altaffiltext{7}{\scriptsize Jet Propulsion Laboratory, California Institute of Technology, 4800 Oak Grove Drive, Pasadena, CA 91109, USA}
\altaffiltext{8}{\scriptsize NASA Exoplanet Science Institute, California Institute of Technology, Mail Code 100-22, 770 South Wilson Avenue, Pasadena, CA 91125, USA}
\altaffiltext{9}{\scriptsize Bay Area Environmental Research Institute, 625 2nd St., Ste 209, Petaluma, CA 94952, USA}
\altaffiltext{10}{\scriptsize Department of Astronomy, Yale University, 52 Hillhouse Avenue, New Haven, CT 06511, USA}
\altaffiltext{11}{\scriptsize Wyle Laboratories, 1960 East Grand Ave, Suite 900,  El Segundo, CA 90245, USA}
\altaffiltext{12}{\scriptsize Orbital Sciences Corporation, 2401 E. El Segundo Blvd, Suite 200, El Segundo, CA 90245, USA}
\altaffiltext{13}{\scriptsize Sydney Institute for Astronomy (SIfA), School of Physics, University of Sydney, NSW 2006, Australia}
\altaffiltext{14}{\scriptsize Department of Physics and Astronomy, Aarhus University, Ny Munkegade 120, DK-8000 Aarhus C, Denmark}
\altaffiltext{15}{\scriptsize Pennsylvania State University, 403 Davey Lab, University Park, PA 16802, USA}

\begin{abstract}
We present the seventh \kepler{} planet candidate catalog, which is the first to be based on the entire, uniformly processed, 48 month \kepler{} dataset. This is the first fully automated catalog, employing robotic vetting procedures to uniformly evaluate every periodic signal detected by the Q1--Q17 Data Release 24 (DR24) \kepler{} pipeline. While we prioritize uniform vetting over the absolute correctness of individual objects, we find that our robotic vetting is overall comparable to, and in most cases is superior to, the human vetting procedures employed by past catalogs. This catalog is the first to utilize artificial transit injection to evaluate the performance of our vetting procedures and quantify potential biases, which are essential for accurate computation of planetary occurrence rates. With respect to the cumulative \kepler{} Object of Interest (KOI) catalog, we designate \nnewkois{} new KOIs, of which \nnewpcs{} are dispositioned as planet candidates (PCs). Also, \nfptopc{} KOIs dispositioned as false positives (FPs) in previous \kepler{} catalogs have their disposition changed to PC and \npctofp{} PCs have their disposition changed to FP. This brings the total number of known KOIs to \ntotkois{} and PCs to \ntotpcs{}. We compare the Q1--Q17~DR24 KOI catalog to previous KOI catalogs, as well as ancillary \kepler{} catalogs, finding good agreement between them. We highlight new PCs that are both potentially rocky and potentially in the habitable zone of their host stars, many of which orbit solar-type stars. This work represents significant progress in accurately determining the fraction of Earth-size planets in the habitable zone of Sun-like stars. The full catalog is publicly available at the NASA Exoplanet Archive.
\end{abstract}


\section{Introduction}
\label{introsec}

The \kepler{} instrument is a 0.95 meter aperture, optical (423--897 nm at $>$5\% throughput), space-based telescope that employs 42 CCDs to photometrically observe $\sim$170,000 stars over a field of view of 115 square degrees \citep{Koch2010}. It achieves a combined (intrinsic and instrumental) noise on 12th magnitude solar-type stars of $\sim$30 ppm \citep{Gilliland2011,Christiansen2012} on a 6-hour time-scale. The primary objective of the \kepler{} mission is to determine the frequency of Earth-size planets in the habitable zone around Solar-like stars \citep{Borucki2010a} by searching for the periodic drops in brightness that occur when planets transit their host stars. Observations of the original \kepler{} field lasted from 2009 May 13 until 2013 May 11, when the second of four on-board reaction wheels failed. The spacecraft could no longer maintain the required pointing precision in the original \kepler{} field and was re-purposed for an ecliptic plane mission \citep[K2;][]{Howell2014}. In this paper we focus exclusively on data collected from the original \kepler{} field (19h 22m 40s, +44$\degr$ 30$\arcmin$ 00$\arcsec$).

A series of previously published \kepler{} catalog papers presented an increasingly larger number of planet candidates as additional observations were collected by the spacecraft \citep{Borucki2011a,Borucki2011b,Batalha2013,Burke2014,Rowe2015a,Mullally2015a}. These catalogs have been used extensively in the investigation of planetary occurrence rates \citep[e.g.,][]{Catanzarite2011,Youdin2011,Howard2012,Dressing2013,Fressin2013,Dong2013,Petigura2013b,Foreman-Mackey2014,Mulders2015,Burke2015}, determination of exoplanet atmospheric properties \citep[e.g.,][]{Coughlin2012,Esteves2013,Demory2014,Sheets2014}, and development of planetary confirmation techniques via supplemental analysis and follow-up observations \citep[e.g.,][]{Moorhead2011,Morton2011,Steffen2012,Ford2012,Fabrycky2012,Santerne2012,Adams2012,Colon2012,Adams2013,Barrado2013,Law2014,Lillo-Box2014,Muirhead2014,Plavchan2014,Rowe2014,Dressing2014,Everett2015}. Furthermore, astrophysically variable systems not due to transiting planets have yielded valuable new science on stellar binaries, including eclipsing \citep[e.g.,][]{Prsa2011,Slawson2011,Coughlin2011}, self-lensing \citep{Kruse2014}, beaming \citep{Faigler2011,Shporer2011}, and tidally interacting systems \citep[e.g.,][]{Thompson2012}. While widely used, these previous catalogs involved a substantial amount of manual vetting by a dedicated team of scientists, and as a result were non-uniform (i.e., not every signal was vetted, and those examined were not vetted to the same standard.)

This paper describes the use of a robotic vetting procedure to produce, for the first time, a fully automated and uniform planetary catalog based on the entire \kepler{} mission dataset (Q1--Q17; 48 months; data release 24). This procedure and resulting catalog enables a more accurate determination of planetary occurrence rates, as any potential biases of the robotic vetting can be quantified via artificial transit injection and other tests. However, we note that due to a subtle flaw in the implementation of a veto in the \kepler{} pipeline, a non-uniform planet search was conducted, and thus care should be taken if using this catalog to compute planetary occurrence rates (see \S\ref{falsenegsec}).

In \S\ref{tcesec} we discuss the population of signals possibly due to transiting planets that are identified by the \kepler{} pipeline and used in this catalog. In \S\ref{robosec} we describe the robotic procedure employed to vet and disposition every signal. In \S\ref{resultssec} we list the inputs to and results of the robotic vetting, describe the designation of Kepler Objects of Interest, and explain the subsequent transit-model fitting. In \S\ref{analsec} we compare this catalog to previous and ancillary catalogs, assess the performance of the robotic vetting utilizing the results of artificial transit injection, and highlight and scrutinize new planet candidates that are potentially rocky and in the habitable zone of their host stars. In \S\ref{discusssec} we discuss the scientific impact of this catalog, and what work can be done to further improve and characterize our vetting procedures for the next \kepler{} catalog. Finally, we note that due to the significant number of acronyms that are inherent to any large mission like \kepler{}, in Appendix~\ref{acroappendsec} we list and define all the acronyms used in this paper.

\section{Q1--Q17~DR24 TCE\MakeLowercase{s}}
\label{tcesec}

This catalog is based on \keplers{} 24\th{} data release (DR24), which includes the processing of all data utilizing version 9.2 of the \kepler{} pipeline \citep{Jenkins2010}. This marks the first time that all of the \kepler{} mission data have been processed consistently with the same version of the \kepler{} pipeline. Over a period of 48 months (2009 May 13 to 2013 May 11), subdivided into 17 quarters (Q1--Q17), a total of 198,646 targets were observed, with 112,001 targets observed in every quarter and 86,645 observed in a subset of the 17 quarters \citep{Seader2015}.  The calibrated pixel-level images and processed light curves are publicly available at the Mikulksi Archive for Space Telescopes (MAST)\footnote{http://archive.stsci.edu/kepler}, along with thorough documentation via the \kepler{} Instrument Handbook \citep{VanCleve2009}, the \kepler{} Data Characteristics Handbook \citep{Christiansen2013a}, the \kepler{} Archive Manual \citep{Thompson2014}, and the \kepler{} Data Release 24 Notes \citep{Thompson2015a}.

\citet{Seader2015} discuss in detail the process of identifying threshold crossing events (TCEs), which are periodic flux decrements that may be consistent with the signals produced by transiting exoplanets. Each TCE has an associated \kepler{} Input Catalog (KIC) ID, period, epoch, depth, and duration. For DR24, \citet{Seader2015} identified a total of \ntces{} TCEs, which are publicly available at the NASA Exoplanet Archive\footnote{http://exoplanetarchive.ipac.caltech.edu} in the Q1--Q17~DR24 TCE table. We employ these \ntces{} TCEs as our starting point to produce a planet candidate catalog, with the goal of designating each TCE as a planet candidate (PC) or false positive (FP). In the next two subsections we explore the TCE false positive population (TCEs that are not due to transiting planets) and the false negative population (transiting planets that were not detected).

\subsection{The TCE False Positive Population}

\begin{figure*}
\centering
\includegraphics[width=\linewidth]{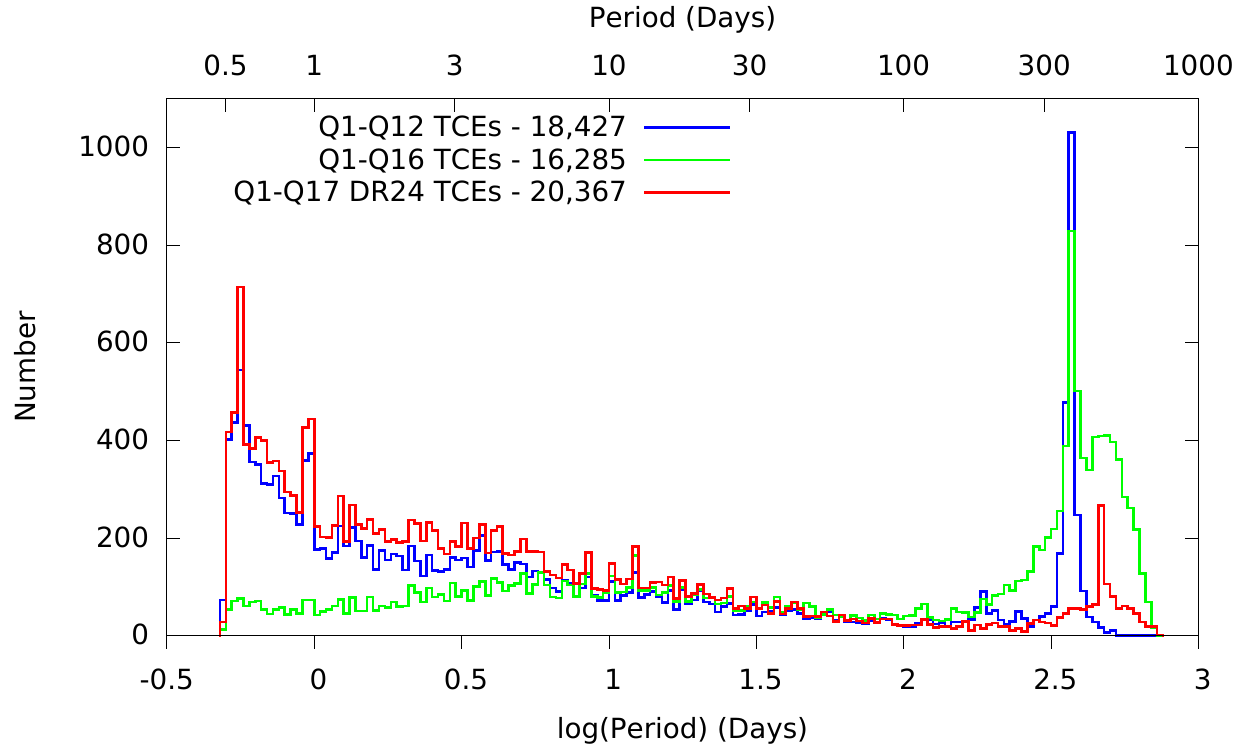}
\caption{The distribution of Threshold Crossing Events (TCEs) as a function of period, with uniform bins in log space. TCEs from Q1--Q12 \citep{Tenenbaum2013} are plotted in blue, TCEs from Q1--Q16 \citep{Tenenbaum2014} are plotted in green, and TCEs from Q1--Q17~DR24 \citep{Seader2015} are plotted in red.}
\label{tce-comp-fig}
\end{figure*}

In Figure~\ref{tce-comp-fig} we plot a histogram of the number of Q1--Q17~DR24 TCEs identified as a function of period \citep{Seader2015}. We also plot the TCE populations from the two previously published searches, which used data from Q1--Q16 \citep{Tenenbaum2014} and Q1--Q12 \citep{Tenenbaum2013}, processed by previous versions of the \kepler{} pipeline. Given that the observed period distribution of transiting planets is thought to be relatively flat and smooth in log space \citep{Howard2012,Fressin2013}, and that the population of TCEs has varied significantly between successive data releases and pipeline versions, it is clear that all of these TCE catalogs contain a large number of false positives.

For Q1--Q17~DR24 and Q1--Q12 there is a particularly large excess at short periods principally due to short-period, quasi-sinusoidal variable stars, e.g., rapid rotators with strong starspots and pulsating stars, as well as eclipsing binary (EB) stars. Spikes seen in this short-period regime are due to contamination from bright variable stars \citep{Coughlin2014a}, such as RR~Lyrae at 0.567 days (-0.25 in log-space), V2083~Cyg at 0.934 days (-0.03 in log-space), and V380~Cyg at 12.426 days (1.09 in log-space). For Q1--Q12 and Q1--Q16 there is a large spike of excess TCEs at $\sim$372 days (2.57 in log-space), which is due to quasi-sinusoidal-like red noise produced by ``rolling-band'' instrumental artifacts \citep{VanCleve2009} that repeat at \keplers{} orbital period. For Q1--Q16, and to a lesser extent Q1--Q17, there is a broad excess of long-period TCEs at periods $\gtrsim$200 days. These are due to short-duration systematics (caused by cosmic rays, flares, star spots, stellar pulsation, edge effects around gaps, and similar features) that occur throughout the light curves, and produce a TCE when three events happen to be equally spaced in time. In the Q1--Q17 data, a spike at $\sim$459 days (2.66 in log-space) can be seen, corresponding to TCEs that were generated due to edge effects from three equally spaced data gaps, and thus this $\sim$459 day systematic is common to many stars across the entire field.

\subsection{The TCE False Negative Population}
\label{falsenegsec}

The injection of artificial transits into the pixel-level data is crucial to fully characterize the false negative rate and compute accurate occurrence rates. The completeness (i.e., how often a transiting planet signal is recovered) of the \kepler{} pipeline has been measured for both individual transit events \citep{Christiansen2013b} and multiple transit events spanning a year of data \citep{Christiansen2015b}. An injection run for the entire Q1--Q17~DR24 dataset has been completed and the results are publicly available \citep{Christiansen2015a}. We employ these results in quantifying the accuracy of our planet candidate catalog (see \S\ref{injectsec}).

In the Q1--Q17~DR24 version of the \kepler{} pipeline, a new veto was added called the ``statistical bootstrap test'', which adjusts the detection threshold of the pipeline to account for the presence of non-Gaussian noise --- see the appendices of \citet{Jenkins2002b}, \citet{Seader2015}, and \citet{Jenkins2015} for details. While this test was successful in eliminating many long-period false positives compared to the previous Q1--Q12 and Q1--Q16 runs (see Figure~\ref{tce-comp-fig}), a subtle flaw introduced excess noise into the statistic. This eliminated a significant number of valid long-period, transit-like signals, especially at low signal-to-noise (SNR), which possibly included some previously designated near Earth-size planet candidates in the habitable zones of their host stars from \citet{Rowe2015a} and \citet{Mullally2015a}. Results from the Q1--Q17~DR24 transit injection run \citep{Christiansen2015a} also indicate that the bootstrap test introduced a period-dependent, non-uniform planet search, which complicates the computation of planetary occurrence rates. Future \kepler{} pipeline runs will not employ the bootstrap test as a veto within the transiting planet search (TPS) module, but rather retain a correctly implemented version as a diagnostic metric. 

While not true false negatives, as they are not transiting planets, we also note that on-target, contact eclipsing binary candidates identified by the \kepler{} Eclipsing Binary Working Group\footnote{http://keplerebs.villanova.edu} (EBWG) \citep{Prsa2011,Slawson2011,Kirk2015} were purposely excluded from this transit search, as sinusoidal and quasi-sinusoidal signals are not considered to be transit-like for mission purposes, and significantly increase processing time. There was a total of \nebexclude{} targets excluded, which we list in Table~\ref{ebexcludetab}. ``Contact'' is defined as having a morphology parameter \citep{Matijevic2012} greater than 0.6. Detached eclipsing binaries were not excluded as they are sufficiently transit-like to include in this catalog. Stars that were not searched for transits can also be identified by lacking a value for ``duty cycle'' in the Q1--Q17~DR24 stellar table, which is publicly available at the NASA Exoplanet Archive.

\begin{deluxetable}{c}
\tablecolumns{1}
\tablewidth{\linewidth}
\tablecaption{The \nebexclude{} Contact Eclipsing Binaries Excluded from the Q1--Q17~DR24 \kepler{} Pipeline Transit Search}
\tablehead{\colhead{KIC ID}}
001433410\\
001572353\\
001573836\\
001868650\\
002012362\\
002141697\\
002159783\\
002162283\\
002302092\\
002305277\\
\nodata
\enddata
\tablecomments{Table~\ref{ebexcludetab} is published in its entirety in the electronic edition of the Astrophysical Journal. A portion is shown here for guidance regarding its form and content.}
\label{ebexcludetab}
\end{deluxetable}

\section{Robotic Vetting}
\label{robosec}

In previous planet candidate catalogs \citep{Borucki2011a,Borucki2011b,Batalha2013,Burke2014,Rowe2015a}, various plots and diagnostics for each TCE were visually examined by members of the Threshold Crossing Event Review Team (TCERT), which consists of professional scientists who have a thorough understanding of \kepler{} data systematics and the various types of false positive scenarios. \citet{Mullally2015a} employed partial automation in the Q1--Q16 catalog through the use of three simple parameter cuts, principally to cull out a large number of long-period false positives, as well as a robotic procedure to identify a particular subset of centroid offsets \citep[see \S5.2 of][]{Mullally2015a}.

The need to fully automate the dispositioning of TCEs, a long-standing objective of the \kepler{} mission, is principally driven by the desire to compute accurate planet occurrence rates, which requires that every TCE be dispositioned in a uniform manner so that it can be subjected to quantitative evaluation. As manual inspection by TCERT members is very time-consuming, it is often not feasible to examine each of the $\sim$20,000 TCEs produced by the \kepler{} pipeline. While TCERT members are well-trained, as humans they do not always agree with each other, and individuals may disposition a given TCE differently depending on external factors such as the time of day, their mood, other TCEs examined recently, etc. However, humans are naturally adept at pattern recognition and categorization, and TCERT has developed an efficient and comprehensible workflow procedure, based on understood physical processes, while working on the previous six planet candidate catalogs. 

Thus, for automating the TCE dispositioning process, we have specifically chosen a robotic vetting procedure that operates via a series of simple decision trees. Hereafter referred to as the ``robovetter'', it attempts to mimic the well-known human vetting process, providing a specific reason for dispositioning any TCE as a false positive. The robovetter was initially developed based on the results of the Q1--Q16 catalog \citep{Mullally2015a} and then further refined based on the results of manual checks on the the Q1--Q17~DR24 dataset by TCERT members.

In \citet{Rowe2015a} and \citep{Mullally2015a}, FP TCEs were assigned to one or more of the following false positive categories:

\begin{itemize}
  \item ``Not Transit-Like'': a TCE whose light curve is not consistent with that of a transiting planet or eclipsing binary, such as instrumental artifacts and non-eclipsing variable stars.
  \item ``Significant Secondary'': a TCE that is observed to have a significant secondary event, indicating that the transit-like event is most likely caused by an eclipsing binary. (Self-luminous, hot Jupiters with a visible secondary eclipse are also in this category, but are still given a disposition of PC.)
  \item ``Centroid Offset'': a TCE whose signal is observed to originate on a nearby star, rather than the target star, based on examination of the pixel-level data.
  \item ``Ephemeris Match Indicates Contamination'': a TCE that has the same period and epoch as another object, and is not the true source of the signal given the relative magnitudes, locations, and signal amplitudes of the two objects.
\end{itemize}

\begin{figure*}
\centering
\includegraphics[width=\linewidth]{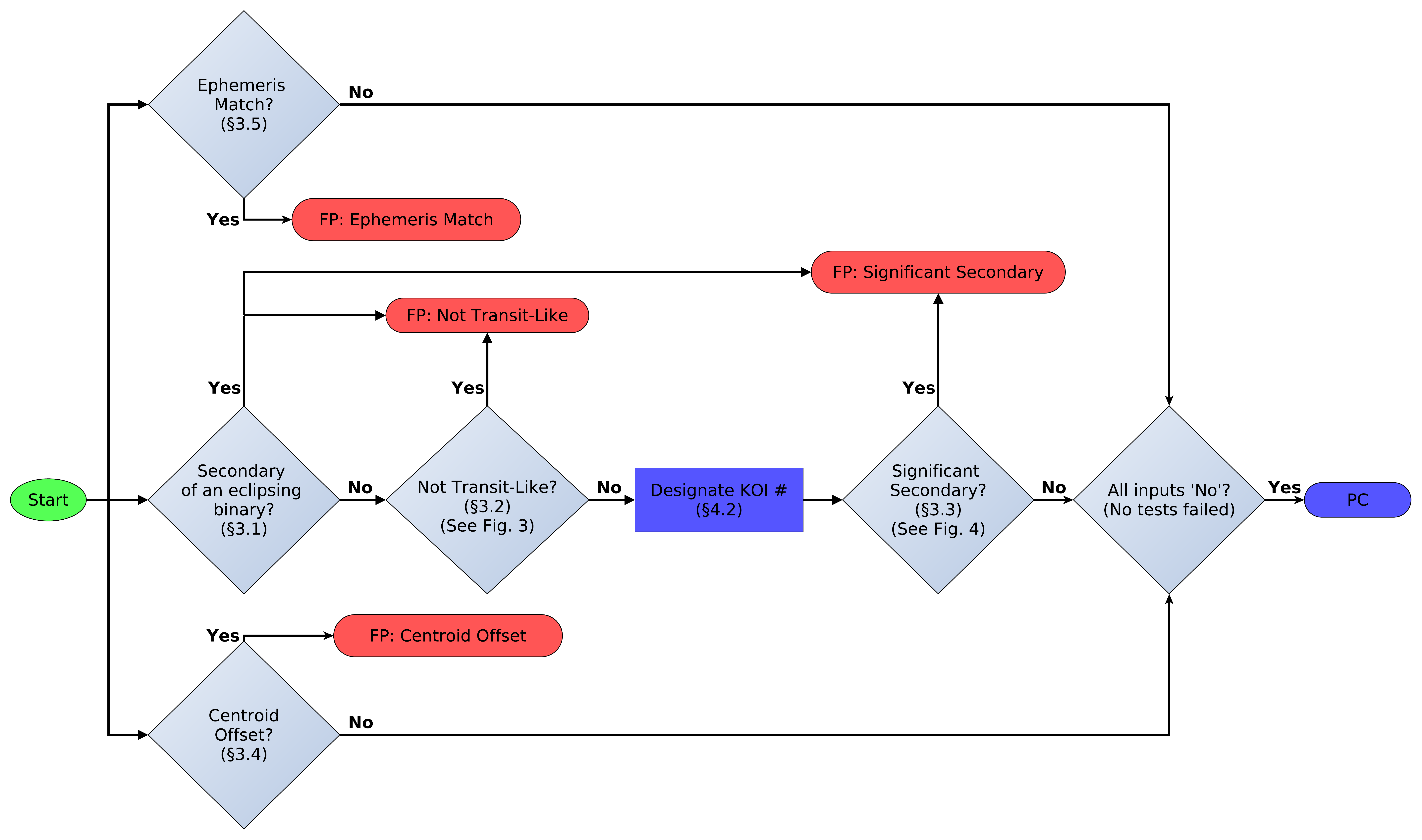}
\caption{Overview flowchart of the robovetter. Diamonds represent ``yes'' or ``no'' decisions that are made with quantitative metrics. A TCE is dispositioned as a FP if it fails any test (a ``yes'' decision) and is placed in one or more of the FP categories. If a TCE passes all tests (a ``no'' decision for all tests) it is dispositioned as a PC. The section numbers on each component correspond to the sections in this paper where these tests are discussed. More in-depth flowcharts are provided for the not transit-like and significant secondary modules in Figures~\ref{robovetter-transitlike-fig} and \ref{robovetter-sigsec-fig}.}
\label{robovetter-overview-fig}
\end{figure*}

\noindent In Figure~\ref{robovetter-overview-fig} we present a flowchart that outlines our robotic vetting procedure. As can be seen, each TCE is subjected to a series of ``yes'' or ``no'' questions (represented by diamonds) that either disposition it into one or more of the four FP categories, or else disposition it as a PC. Behind each question is a series of more specific questions, each answered by quantitative tests. These tests are designed with the same ``innocent until proven guilty'' approach that was used by TCERT members in previous catalogs, such that no TCE is dispositioned as a FP without substantial evidence. Quantitatively we are aiming to preserve at least $\sim$95\% of injected transits while rejecting as many false positives as possible.

We note that for all of the robovetter tests that require a phased light curve and model fit, we utilize two different detrendings and model fits. In the \kepler{} pipeline, the Data Validation (DV) module produces a harmonic-removed, median-detrended, phased flux light curve, along with a transit model fit \citep{Wu2010}. However, the harmonic remover is known to suppress short-period ($\lesssim$ 3 days) signals such that short-period eclipsing binaries with visible secondaries can appear as transiting planets with no visible secondary \citep{Christiansen2013b}. It can also make variable stars with semi-coherent variability, such as starspots or pulsations, appear as transit-like signals. Thus, we create phased flux light curves via an alternate detrending method that utilizes the pre-search data conditioned (PDC) time-series light curves and the non-parametric penalized least squares detrending method of \citet{Garcia2010}, which includes only the out-of-transit points when computing the filter. These alternately detrended light curves are then phased and fit with a simple trapezoidal transit model. This alternate detrending technique is effective at accurately detrending short-period eclipsing binaries and variable stars, i.e., preserving their astrophysical signal. Every test that is applied to the DV phased light curves is also applied to the alternate detrending --- failing a test using either detrending results in the TCE being classified as a FP.

The robovetter first checks if the TCE corresponds to a secondary eclipse associated with an already examined system. If not, the robovetter then checks if the TCE is transit-like or not. If it is transit-like, the robovetter then looks for the presence of a secondary eclipse. In parallel, the robovetter also looks for evidence of a centroid offset and an ephemeris match to other TCEs and variable stars in the \kepler{} field. In the following subsections we describe in detail each of these tests in the order in which they are performed by the robovetter.

\subsection{The TCE is the Secondary of an Eclipsing Binary}

If a TCE under examination is not the first one in a system, the robovetter checks if there exists a previous TCE with a similar period that was designated as a FP due to a significant secondary (see~\S\ref{sigsecsec}). To compute whether two TCEs have the same period within a given statistical threshold, we employ the period matching criteria of \citet[][see equations 1-3]{Coughlin2014a}, $\sigma_{P}$, where higher values of $\sigma_{P}$ indicate more significant period matches. We re-state the equations here as:

\begin{equation}
\label{peq1}
\Delta P = \frac{P_{A}-P_{B}}{P_{A}}\\
\end{equation}

\begin{equation}
\label{peq2}
\Delta P^{\prime} = \textrm{abs}(\Delta P - \textrm{rint}(\Delta P))\\
\end{equation}

\begin{equation}
\label{peq3}
\sigma_{P} = \sqrt{2}\cdot\textrm{erfcinv}(\Delta P^{\prime})\\
\end{equation}

\noindent where $P_{A}$ is the period of the shorter-period TCE, $P_{B}$ is the period of the longer-period TCE, $rint()$ rounds a number to the nearest integer, $abs()$ yields the absolute value, and erfcinv() is the inverse complementary error function. We consider any value of $\sigma_{P}$ $>$ 3.25 to indicate significantly similar periods.

If the current TCE is (1) in a system that has a previous TCE dispositioned as a FP due to a significant secondary, (2) matches the previous TCE's period with $\sigma_{P}$ $>$ 3.25, and (3) is separated in phase from the previous TCE by at least 2.5 times the transit duration, then the current TCE is considered to be a secondary eclipse. In this case, it is designated as a FP and is classified into both the not transit-like and significant secondary FP categories --- a unique combination that can be used to identify secondary eclipses while still ensuring they are not assigned \kepler{} Object of Interest numbers (see \S\ref{koisec}). Note that since the \kepler{} pipeline identifies TCEs in order of their SNR, from high to low, sometimes a TCE identified as a secondary can have a deeper depth than the primary, depending on their relative durations and shapes.

There are two cases where we modify the three criteria above. First, it is possible that the periods of two TCEs will meet the period matching criteria, but be different enough to have their relative phases shift significantly over the $\sim$4 year mission duration. Thus, the potential secondary TCE is actually required to be separated in phase by at least 2.5 times the previous TCE's transit duration over the entire mission time frame in order to be labeled as a secondary. Second, the \kepler{} pipeline will occasionally detect the secondary eclipse of an EB at a half, third, or some smaller integer fraction of the orbital period of the system, such that the epoch of the detected secondary coincides with that of the primary. Thus, for the non-1:1 period ratio cases, we do not impose the phase separation requirement. (Note that equations~\ref{peq1}-\ref{peq3} allow for integer period ratios.)

\subsection{Not Transit-Like}
\label{nottransitlikesec}

\begin{figure*}
\centering
\includegraphics[width=\linewidth]{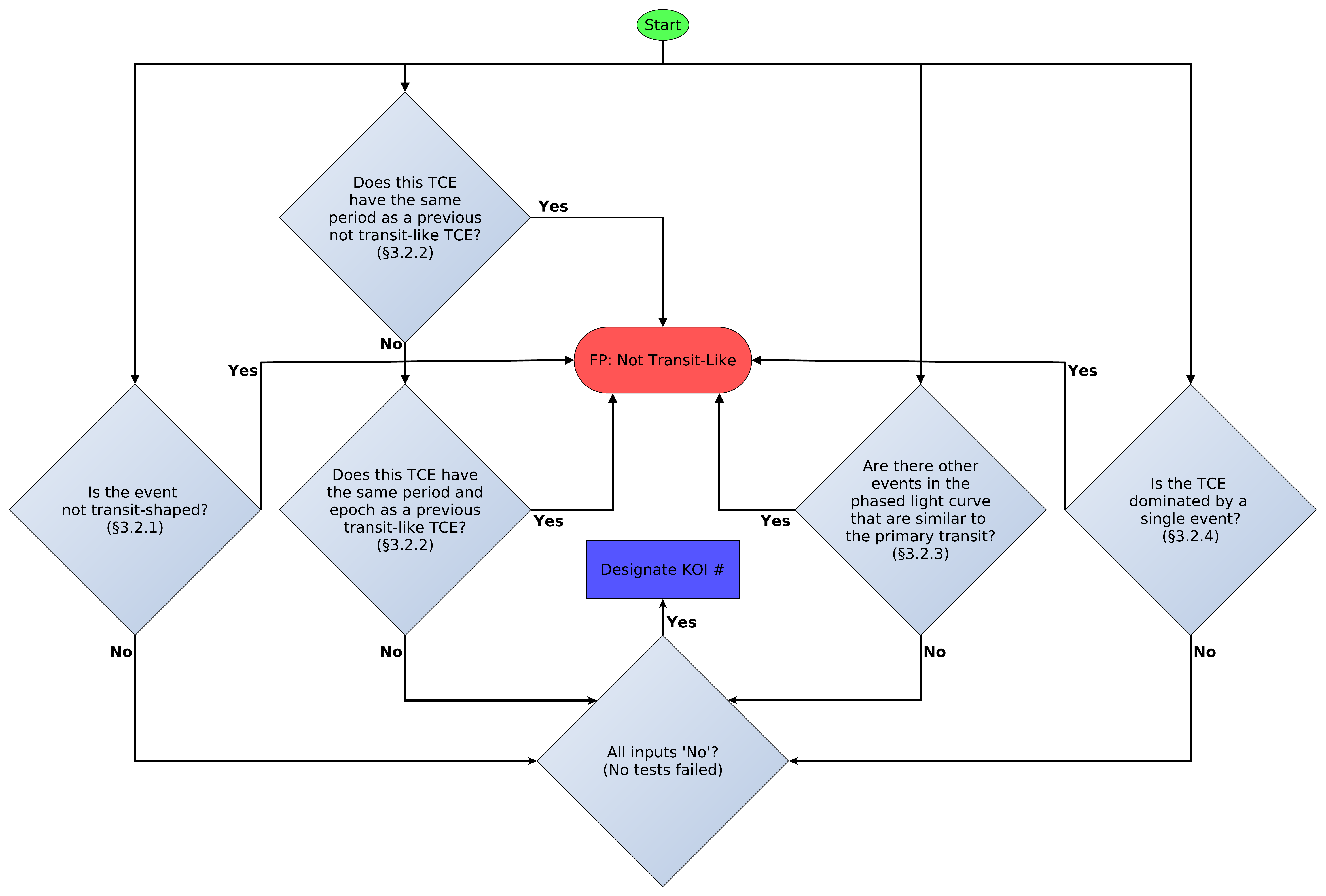}
\caption{Not transit-like flowchart of the robovetter. Diamonds represent ``yes'' or ``no'' decisions that are made with quantitative metrics. If a TCE fails any test (via a ``yes'' response to any decision) then it is dispositioned as a not transit-like FP. If a TCE passes all tests (via a ``no'' response to all decisions), then it is given a KOI number and passed to the significant secondary module (see \S\ref{sigsecsec} and Figure~\ref{robovetter-sigsec-fig}). The section numbers on each decision diamond correspond to the sections in this paper where these tests are discussed.}
\label{robovetter-transitlike-fig}
\end{figure*}

A very large fraction of false positive TCEs have light curves that do not resemble a detached transiting or eclipsing object. These include quasi-sinusoidal light curves from pulsating stars, starspots, and contact binaries, as well as more sporadic light curves due to instrumental artifacts. In previous planet candidate catalogs a process called ``triage'' was employed whereby the human vetters looked at the phased light curves to determine whether the TCEs were not transit-like, or should be given \kepler{} Object of Interest (KOI) numbers, which are used to keep track of transit-like systems over multiple \kepler{} pipeline runs. We thus employ a series of algorithmic tests to reliably identify these not transit-like FP TCEs, as shown by the flowchart in Figure~\ref{robovetter-transitlike-fig}.

\subsubsection{Not Transit-Shaped}

The human members of TCERT were given training and diagnostic plots that allowed them to quickly distinguish between a quasi-sinusoidal shaped light curve and one that is more detached due to a transit or eclipse. Also, they were trained to recognize if an individual event is due to a transit or a systematic feature, such as a sudden discontinuity in the light curve. As such, we sought metrics for the robovetter that would similarly detect quasi-sinusoidal light curves and systematics.

\paragraph{The LPP Metric}
\label{lppsec}

Many short-period false positives are due to variable stars that exhibit a quasi-sinusoidal phased light curve. \citet{Matijevic2012} used a technique known as Local Linear Embedding (LLE), a dimensionality reduction algorithm, to classify the ``detachedness'' of \kepler{} eclipsing binary light curves on a scale of 0 to 1, where 0 represented fully detached systems with well-separated, narrow eclipses and 1 represented contact binaries with completely sinusoidal light curves. We use a similar technique, known as Local Preserving Projections \citep[][LPP]{He2004}, to distinguish transit-like signals from not transit-like signals \citep{Thompson2015b}. LPP returns a single number that represents the similarity of a TCE's shape to that of known transits. Unlike LLE, LPP can be applied to any TCE, not just those that lie within the parameter space of the training set. Thus, LPP is more suitable for separating transit-like TCEs from all other not transit-like TCEs, and can be run on artificially injected transits.

To calculate the LPP metric we start with detrended \kepler{} light curves. We then fold and bin each light curve into 141 points, ensuring adequate coverage of both the in- and out-of-transit portions of the light curve. We exclude points near a phase of 0.5, as the presence of a secondary eclipse in a short-period binary may unduly influence the LPP value, and we seek to classify detached eclipsing binaries as transit-like. These 141 points act as the initial number of dimensions that describe each TCE. Using a subset of known transit-like TCEs, we create a map from the initial 141 dimensions down to 20 dimensions. We apply this map to all TCEs and measure the average Euclidean distance of each to the 15 nearest known transit-like TCEs. This average distance is the value of the LPP metric. When small, it means other transit-like TCEs are nearby in the 20 dimensional space and thus is likely to be shaped like a transit. We calculate this LPP transit metric for all TCEs using both the DV and the alternate detrending, as described in \S\ref{nottransitlikesec}.

In order to quantitatively determine a threshold between transit-like and not transit-like, we run the LPP classifier on both detrendings of the injected transits (see \S\ref{injectsec}), which we know a priori are all transit-shaped, barring any light curve distortion due to detrending. We then fit a Gaussian to the resulting distribution, computing its median and standard deviation. We then select a maximum LPP cutoff such that we expect less than one false negative in \ntces{} TCEs, via 

\begin{equation}
\sigma_{\rm LPP} = \sqrt{2}\cdot \textrm{erfcinv}(1/N_{\rm TCEs})
\end{equation}

\noindent where $N_{TCEs}$ = \ntces{}, yielding $\sigma_{\rm LPP}$ = 4.06. Any TCE with a LPP value greater than the median plus 4.06 times the standard deviation, using either detrending, is considered not transit-like.

\paragraph{The Marshall Metric}

A number of long-period false positives are a result of three or more systematic events that happen to be equidistant in time and produce a TCE. There are two prominent types of systematic events in \kepler{} data: sudden pixel sensitivity dropouots (SPSDs) and step-wise discontinuities. SPSDs are due to cosmic ray impacts that temporarily reduce the detection sensitivity of the impacted pixels, resulting in a sudden drop in flux followed by an asymptotic rise back to the baseline flux level over a timescale of a few hours \citep{VanCleve2009}. Step-wise discontinuities are sudden jumps in the baseline flux level, in either the positive or negative flux direction, and are typically due to imperfect detrending, but may have other causes. If a TCE is due to several of these events that are of similar SNR, they will not be flagged as false positives without examining the shape of their individual events.

In order to detect TCEs due to SPSDs and step-wise discontinuities, we developed the ``Marshall'' metric \citep{Mullally2015b}. Marshall fits a transit, SPSD, and step-wise discontinuity model to each individual event of a long-period TCE. The Bayesian Information Criterion \citep[BIC;][]{Schwarz1978} is then used to select which model best fits each individual transit event given each model's number of degrees of freedom. If either the SPSD or step-wise discontinuity model have a lower BIC value than the transit model by a value of 10 or more for a given transit event, then that event is determined to be due to a systematic rather than a transit. After evaluating each individual event, if there are fewer than three events that are determined to be due to transits, the TCE is dispositioned as a not transit-like false positive. This is in line with the \kepler{} mission requirement of detecting at least three valid transits in order to generate a TCE.

\subsubsection{Previous TCE With Same Period}

Most quasi-sinusoidal false positives produce multiple TCEs at the same period, or at integer ratios of each other. If a TCE in a system has been declared as not transit-like due to another test, it is logical that all subsequent TCEs in that system at the same period, or ratios thereof, should also be dispositioned not transit-like. Thus, we match the period of a given TCE to all previous not transit-like FPs via equations~\ref{peq1}-\ref{peq3}. If the current TCE has a period match with $\sigma_{P}$ $>$ 3.25 to a prior not transit-like FP, it is also dispositioned as a not transit-like FP.

Similarly, some TCEs are produced that correspond to the edge of a previously identified transit-like TCE in the system. This often results when the previous TCE corresponding to a transit or eclipse is not completely removed prior to searching the light curve for another TCE. Thus, we match the period of a given TCE to all previous transit-like TCEs via equations~\ref{peq1}-\ref{peq3}.  If the current TCE has a period match with $\sigma_{P}$ $>$ 3.25 to a prior transit-like FP, and the two epochs are separated in phase by less than 2.5 transit durations, the current TCE is dispositioned as a not transit-like FP. For clarity, we note that it is sometimes possible that the periods of two TCEs will meet the period matching criteria, but be different enough to have their epochs shift significantly in phase over the $\sim$4 year mission duration. Thus, if they are separated in phase by less than 2.5 transit durations at any point in the mission time frame, the current TCE is dispositioned as a not transit-like FP.

\subsubsection{The Model-Shift Uniqueness Test}
\label{notuniquetcesec}

If a TCE under investigation is truly a PC, there should not be any other transit-like events in the light curve with a depth, duration, and period similar to the primary signal, in either the positive or negative flux directions, i.e., the transit event should be unique in the phased light curve. Many false positives are due to quasi-sinusoidal signals (see \S\ref{tcesec}) and thus are not unique in the phased light curve. In order to identify these cases, TCERT developed a ``model-shift uniqueness test'' and used it extensively for identifying false positives in both the Q1--Q12 \citep{Rowe2015a} and Q1--Q16 \citep{Mullally2015a} planet candidate catalogs.

See \S3.2.2 of \citet{Rowe2015a} and page 20 of \citet{Coughlin2014b} for figures and a detailed explanation of the ``model-shift uniqueness test'', but in brief, after removing outliers, the best-fit model of the primary transit is used as a template to measure the best-fit depth of the transit model at all other phases. The deepest event aside from the primary (pri) transit event is labeled as the secondary (sec) event, the next-deepest event is labeled as the tertiary (ter) event, and the most positive (pos) flux event (i.e., shows a flux brightening) is labeled as the positive event. The significances of these events ($\sigma_{\rm Pri}$, $\sigma_{\rm Sec}$, $\sigma_{\rm Ter}$, and $\sigma_{\rm Pos}$) are computed assuming white noise as determined by the standard deviation of the light curve residuals. Also, the ratio of the red noise (at the timescale of the transit duration) to the white noise ($F_{\rm Red}$) is computed by examining the standard deviation of the best-fit depths at phases outside of the primary and secondary events.  When examining all events among all TCEs, the minimum threshold for an event to be considered statistically significant is given by

\begin{equation}
\sigma_{\rm FA} = \sqrt{2}\cdot \textrm{erfcinv}\left(\frac{T_{\rm dur}}{P \cdot N_{\rm TCEs}}\right)
\end{equation}

\noindent where $T_{\rm dur}$ is the transit duration, and $P$ is the period. (The quantity $P$/$T_{\rm dur}$ represents the number of independent statistical tests for a single target.) When comparing two events from the same TCE, the minimum difference in their significances in order to be considered distinctly different is given by

\begin{equation}
\sigma'_{\rm FA} = \sqrt{2}\cdot \textrm{erfcinv}\left(\frac{T_{\rm dur}}{P}\right)
\end{equation} 

In the robovetter, we disposition a TCE as a not transit-like FP if either ${\sigma_{\rm Pri}/F_{\rm Red} < \sigma_{\rm FA}}$, ${\sigma_{\rm Pri} - \sigma_{\rm Ter} < \sigma'_{\rm FA}}$, or ${\sigma_{\rm Pri} - \sigma_{\rm Pos} < \sigma'_{\rm FA}}$ for either the DV or alternate detrending. These criteria ensure that the primary event is statistically significant when compared to the systematic noise level of the light curve, the tertiary event, and the positive event, respectively.

\subsubsection{Dominated by Single Event}

The depths of individual transits of planet candidates should be equal to each other, and thus assuming constant noise levels, the SNR of individual transits should be nearly equivalent as well. In contrast, most of the long-period FPs that result from three or more equidistant systematic events are dominated in SNR by one of those events. The \kepler{} pipeline measures detection significance via the Multiple Event Statistic (MES), which is calculated by combining the Single Event Statistic (SES) of all the individual events that comprise the TCE --- both the MES and SES are measures of SNR. Assuming all individual events have equal SES values,

\begin{equation}
{\rm MES} = \sqrt{N_{\rm Trans}} \cdot {\rm SES}
\end{equation}

\noindent where $N_{\rm Trans}$ is the number of transit events that comprise the TCE. Thus, SES/MES = 0.577 for a TCE with three transits, and less for a greater number of transits. If the largest SES value of a TCE's transit events, ${\rm SES}_{\rm Max}$, divided by the MES is much larger than 0.577, this indicates that one of the individual events dominates when calculating the SNR.

In the robovetter, for TCEs with periods greater than 90 days, if ${{\rm SES}_{\rm Max} / {\rm MES} > 0.9}$ it is dispositioned as a not transit-like false positive. The value of 0.9 was empirically chosen based on the results of transit injection (\S\ref{injectsec}) to reject a minimal number of valid planetary candidates, accounting for natural deviations of SES values due to light curve systematics and changes in local noise levels. The period cutoff of 90 days is applied because short-period TCEs can have a large number of individual transit events, which dramatically increases the chance of one event coinciding with a large systematic feature, thus producing a large ${{\rm SES}_{\rm Max} / {\rm MES}}$ value despite being a valid planetary signal.

\subsection{Significant Secondary}
\label{sigsecsec}

\begin{figure*}
\centering
\includegraphics[width=\linewidth]{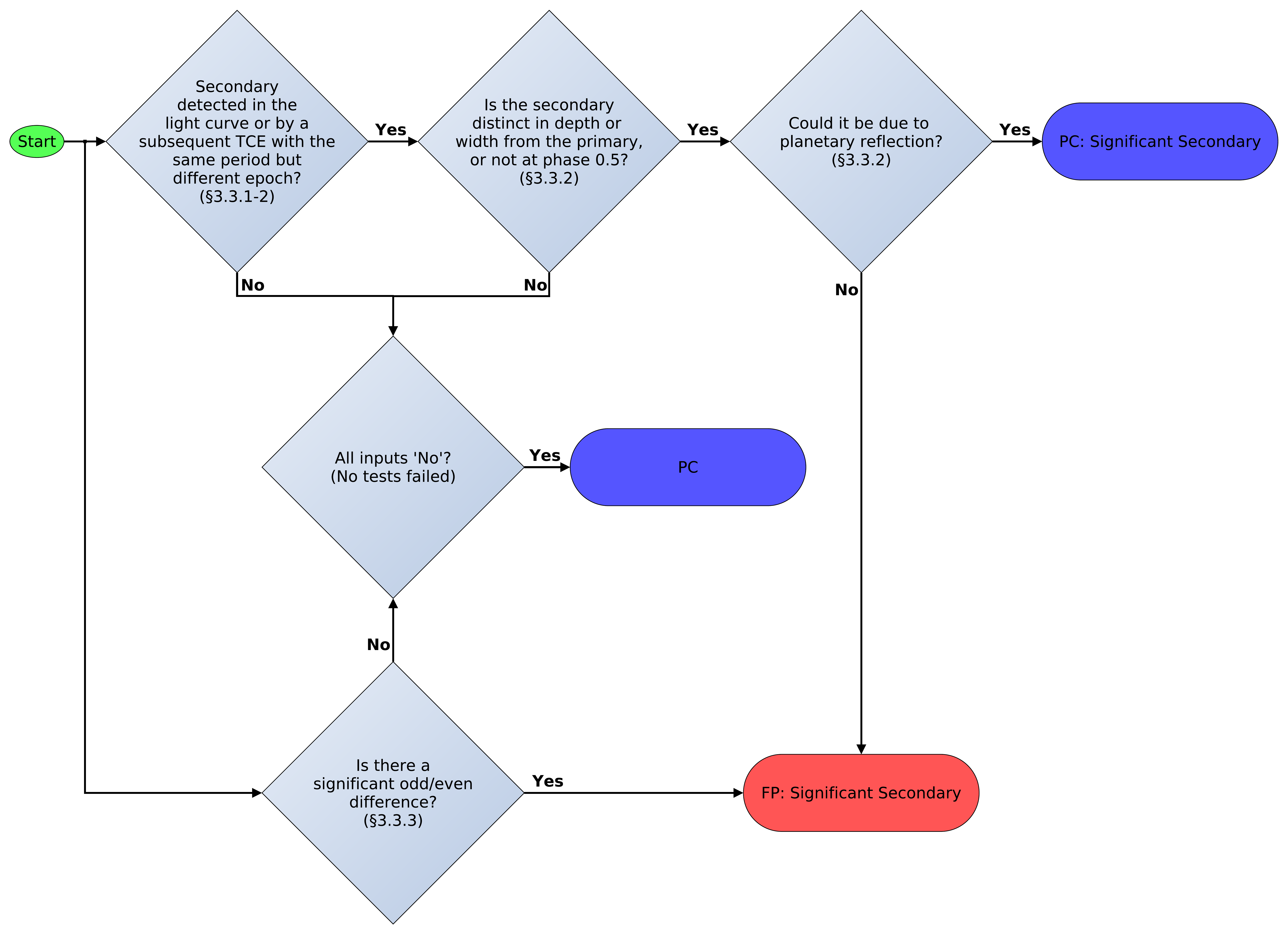}
\caption{Flowchart describing the significant secondary tests of the robovetter. Diamonds represent ``yes'' or ``no'' decisions that are made with quantitative metrics. The multiple arrows originating from ``Start'' represent decisions that are made in parallel.}
\label{robovetter-sigsec-fig}
\end{figure*}

If a TCE is deemed transit-like by passing all of the tests presented in \S\ref{nottransitlikesec} on both detrendings, it is given a KOI number. However, many of these KOIs are FPs due to eclipsing binaries and contamination from nearby variable stars. In order to produce a uniform catalog, we do not designate any TCE a FP on the basis of its transit depth or inferred radius --- see \S7 item 6 of \citet{Mullally2015a} for more detail. Thus, being agnostic to stellar parameters, the only way to definitively detect an EB via a \kepler{} light curve is by detecting a significant secondary eclipse. We employ a series of robotic tests to detect secondary eclipses, as shown by the flowchart in Figure~\ref{robovetter-sigsec-fig}.

\subsubsection{Subsequent TCE With Same Period}

Once the \kepler{} pipeline detects a TCE in a given system, it removes the data corresponding to this event and re-searches the light curve. It is thus able to detect the secondary eclipse of an EB as a subsequent TCE, which will have the same period as, but different epoch than, the primary TCE. Thus, utilizing equations~\ref{peq1}-\ref{peq3}, the robovetter dispositions a TCE as a FP due to a significant secondary if its period matches a subsequent TCE within the utilized tolerance ($\sigma_{P}$ $>$ 3.25) and they are separated in phase by at least 2.5 times the transit duration. For clarity, we note again that it is sometimes possible that the periods of two TCEs will meet the period matching criteria, but be different enough to have their epochs shift significantly in phase over the $\sim$4 year mission duration. Thus, this phase separation requirement is required to be upheld over the entire mission duration in order to disposition the TCE as a FP due to a significant secondary.

Occasionally the \kepler{} pipeline will detect the secondary eclipse of an EB at half, third, or some smaller integer fraction of the orbital period of the system. In these cases, the epoch of the TCE corresponding to the secondary will overlap with that of the primary. These cases are accounted for by not requiring a phase separation of at least 2.5 transit durations when a period ratio other than unity is detected. (Note that equations~\ref{peq1}-\ref{peq3} allow for integer period ratios.) While this approach will likely classify any multi-planet system in an exact 2:1 orbital resonance as a FP due to a significant secondary, in practice this is non-existent. Exact 2:1 orbital resonances, where ``exact'' means the period ratio is close enough to 2.0 over the $\sim$4 year mission duration to avoid any drift in relative epoch, appear to be extremely rare \citep{Fabrycky2014}. Also, they would produce strong transit timing variations, which would likely preclude their detection. The \kepler{} pipeline employs a strictly linear ephemeris when searching for TCEs, and thus while planets with mild transit timing variations (TTVs), e.g., deviations from a linear ephemeris less than the transit duration, are often detected, planets with strong TTVs, e.g., deviations from a linear ephemeris greater than the transit duration, are often not detected.

\subsubsection{Secondary Detected in Light Curve}
\label{secdetectsec}

There are many cases when a secondary eclipse does not produce its own TCE, most often when its MES is below the \kepler{} pipeline detection threshold of 7.1. The model-shift uniqueness test, discussed in \S\ref{notuniquetcesec}, is well-suited to automatically detect secondary eclipses in the phased light curve, as it searches for the next two deepest events aside from the primary event. It is thus able to detect the best-candidate secondary eclipse in the light curve and assess its significance. The robovetter dispositions any TCE as a FP due to a significant secondary if all three of the following conditions are met, for either the DV or alternate detrending: ${\sigma_{\rm Sec}/F_{\rm Red} > \sigma_{\rm FA}}$, ${\sigma_{\rm Sec} - \sigma_{\rm Ter} > \sigma'_{\rm FA}}$, and ${\sigma_{\rm Sec} - \sigma_{\rm Pos} > \sigma'_{\rm FA}}$ (see \S\ref{notuniquetcesec}). These criteria ensure that the secondary event is statistically significant when compared to the systematic noise level of the light curve, the tertiary event, and the positive event, respectively.

There are two exceptions when the above-mentioned conditions are met, but the robovetter does not designate the TCE a false positive. First, if the primary and secondary are statistically indistinguishable, and the secondary is located at phase 0.5, then it is possible that the TCE is a PC that has been detected at twice the true orbital period. Thus, the robovetter labels a TCE with a significant secondary as a PC when ${\sigma_{\rm Pri} - \sigma_{\rm Sec} < \sigma'_{\rm FA}}$ and the phase of the secondary is within 1/4 of the primary transit's duration of phase 0.5. Second, hot Jupiter PCs can have detectable secondary eclipses due to planetary occultations via reflected light and thermal emission \citep{Coughlin2012}. Thus, a TCE with a detected significant secondary is labeled as a PC with the significant secondary flag (in order to facilitate the identification of hot Jupiter occultations) when the geometric albedo is less than 1.0, the planetary radius is less than 30~\re{}, the depth of the secondary is less than 10\% of the primary, and the impact parameter is less than 0.95. The additional criteria beyond the albedo criterion are needed to ensure that this test is only applied to potentially valid planets and not grazing eclipsing binaries. We calculate the geometric albedo by using the stellar mass, radius, and effective temperature from \citet{Huber2014a}, and the values of the period and radius ratio from the DV module of the \kepler{} pipeline.

\subsubsection{Odd/Even Depth Difference}

If the primary and secondary eclipses of an EB are similar in depth, and the secondary is located near phase 0.5, the \kepler{} pipeline may detect them as a single TCE at half the true orbital period of the EB. In these cases, if the primary and secondary depths are dissimilar enough, it is possible to detect it as a FP by comparing the depths of the odd- and even-numbered transit events. Thus, we compute the following statistic, for both the DV and alternate detrending,

\begin{equation}
\sigma_{\rm OE} = \frac{d_{\rm odd} - d_{\rm even}}{\sqrt{\sigma_{odd}^{2} + \sigma_{even}^{2}}} 
\end{equation}

\noindent where $d_{\rm odd}$ is the median depth of the odd-numbered transits, $d_{\rm even}$ is the median depth of the even-numbered transits, $\sigma_{odd}$ is the standard deviation of the depths of the odd-numbered transits, and $\sigma_{even}$ is the standard deviation of the depths of the even-numbered transits. For the alternate detrending with a trapezoidal fit, we use all points that lie within $\pm$30 minutes of the central time of transit, as well as any other points within the in-transit flat portion of the trapezoidal fit. For the DV detrending, we use all points within $\pm$30 minutes of the central time of transit. (This threshold corresponds to the long-cadence integration time of the \kepler{} spacecraft. Including points farther away from the central time of transit degrades the accuracy and precision of the test.) If $\sigma_{\rm OE}$ $>$ 1.7 for either the DV or alternate detrending then the TCE is labeled as a FP due to a significant secondary. The value of 1.7 was empirically derived utilizing manual checks and transit injection.

\subsection{Centroid Offset}

Given that \keplers{} pixels are 3.98\arcsec{} square \citep{Koch2010}, and the typical photometric aperture has a radius of 4--7 pixels \citep{Bryson2010b}, it is quite common for a given target star to be contaminated by light from another star. If that other star is variable, then that variability will be visible in the target aperture at a reduced amplitude. If the variability due to contamination results in a TCE, then it is a false positive, whether the contaminator is an eclipsing binary, planet, or other type of variable star \citep{Bryson2013}. For example, if a transit or an eclipse occurs on a bright star, a shallower event will be observed on a nearby, fainter star. Similarly, a star can be mistakenly identified as experiencing a shallow transit if a deep eclipse occurs on a fainter, nearby source.

The DV module of the \kepler{} pipeline produces ``difference images'' for each quarter, which are made by subtracting the average flux in each pixel during each transit from the flux in each pixel just before and after each transit \citep{Bryson2013}. If the resulting difference image shows significant flux change at a location (centroid) other than the target, then the TCE is likely a FP due to a centroid offset. In prior catalogs, TCERT members manually examined the difference images to look for evidence of a centroid offset, as fully described in \citet{Bryson2013} and \S3.2.3--3.2.6 of \citet{Rowe2015a}. In this catalog, the search for centroid offsets was fully robotized and confirmed to reproduce the results earlier catalogs using human vetting \citep{Mullally2015c}.

In our robotic procedure to detect FPs due to centroid offsets, we first check that the difference image for each quarter contains a discernible stellar image and is not dominated by background noise. This is done by searching for at least 3 pixels that are adjacent to each other and brighter than a given threshold, which is set by the noise properties of the image. We use an iterative sigma clipping approach to eliminate bright pixels when calculating the background noise, as the star often dominates the flux budget of a substantial number of pixels in the aperture.

For the difference images that are determined to contain a discernible stellar image, we first search for evidence of contamination from sources that are resolved from the target. Since resolved sources near the edge of the image may not be fully captured, Pixel Response Function (PRF --- \keplers{} point spread function convolved with the image motion and the intra-pixel CCD sensitivity) fitting approaches do not often work well to detect them. Instead, we check if the location of the brightest pixel in the difference image is more than 1.5 pixels from the location of the target star. If at least two-thirds of the quarterly difference images show evidence of an offset by this criterion, we disposition the TCE as a FP due to a centroid offset. Note that FPs due to stars located many pixels from the target, i.e., far outside the target's image, are not detected by this approach, but rather through ephemeris matching (see \S\ref{ephemmatchsec}).

If no centroid offset is identified by the previous method, we then look for contamination from sources that are unresolved from the target. We measure the PRF-fit centroid of the difference images and search for statistically significant shifts with respect to the PRF centroid of both the out-of-transit images, as well as the catalog position of the source. Following \citet{Bryson2013}, a TCE is marked as a FP due to a centroid offset if there is a 3$\sigma$ significant offset larger than 2$\arcsec$, or a 4$\sigma$ offset larger than 1$\arcsec$. \citet{Mullally2015c} show that when simulated transits are injected at the catalog position of \kepler{} stars, these robotic methods result in $<$1\% of valid planet candidates being marked incorrectly as FPs.

Note that if there are less than three difference images with a discernible stellar image, no tests are performed, and the TCE is not declared a FP by the centroid module.

\subsection{Ephemeris Matching}
\label{ephemmatchsec}

Another method for detecting FPs due to contamination is to compare the ephemerides (periods and epochs) of TCEs to each other, as well as other known variable sources in the \kepler{} field. If two targets have the same ephemeris within a specified tolerance, then at least one of them is a FP due to contamination. \citet{Coughlin2014a} used Q1--Q12 data to compare the ephemerides of KOIs to each other and eclipsing binaries known from both \kepler{}- and ground-based observations. They identified over 600 FPs via ephemeris matching, of which over 100 were not known as FPs via other methods. They also identified four main mechanisms of contamination. The results of \citet{Coughlin2014a} were incorporated in \citet[][see \S3.3]{Rowe2015a}. \citet[][see \S5.3]{Mullally2015a} slightly modified the ephemeris matching process of \citet{Coughlin2014a}, and applied it to all of the Q1--Q16 TCEs, as well as known KOIs and EBs, identifying nearly 1,000 TCEs as FPs. 

In this Q1--Q17~DR24 catalog, we use the same method as \citet{Coughlin2014a}, with the modifications of \citet[][see \S5.3]{Mullally2015a}, to match the ephemerides of all Q1--Q17~DR24 TCEs \citep{Seader2015} to the following sources:

\begin{itemize}
 \item Themselves.
 \item The list of \npredrtwentyfourkois{} KOIs from the NASA Exoplanet Archive cumulative KOI table after the closure of the Q1--Q16 table and publication of the last catalog \citep{Mullally2015a}.
 \item The \kepler{} EBWG of \nkebs{} “true” EBs found with \kepler{} data as of 2015 March 11 \citep{Prsa2011,Slawson2011,Kirk2015}.
 \item J.M. Kreiner's up-to-date database of ephemerides of ground-based eclipsing binaries as of 2015 March 11 \citep{Kreiner2004}.
 \item Ground-based eclipsing binaries found via the TrES survey \citep{Devor2008a}.
 \item The General Catalog of Variable Stars \citep[GCVS][]{Samus2015} list of all known ground-based variable stars, published 2015 February 06.
\end{itemize}

\begin{figure*}
\centering
\includegraphics[width=\linewidth]{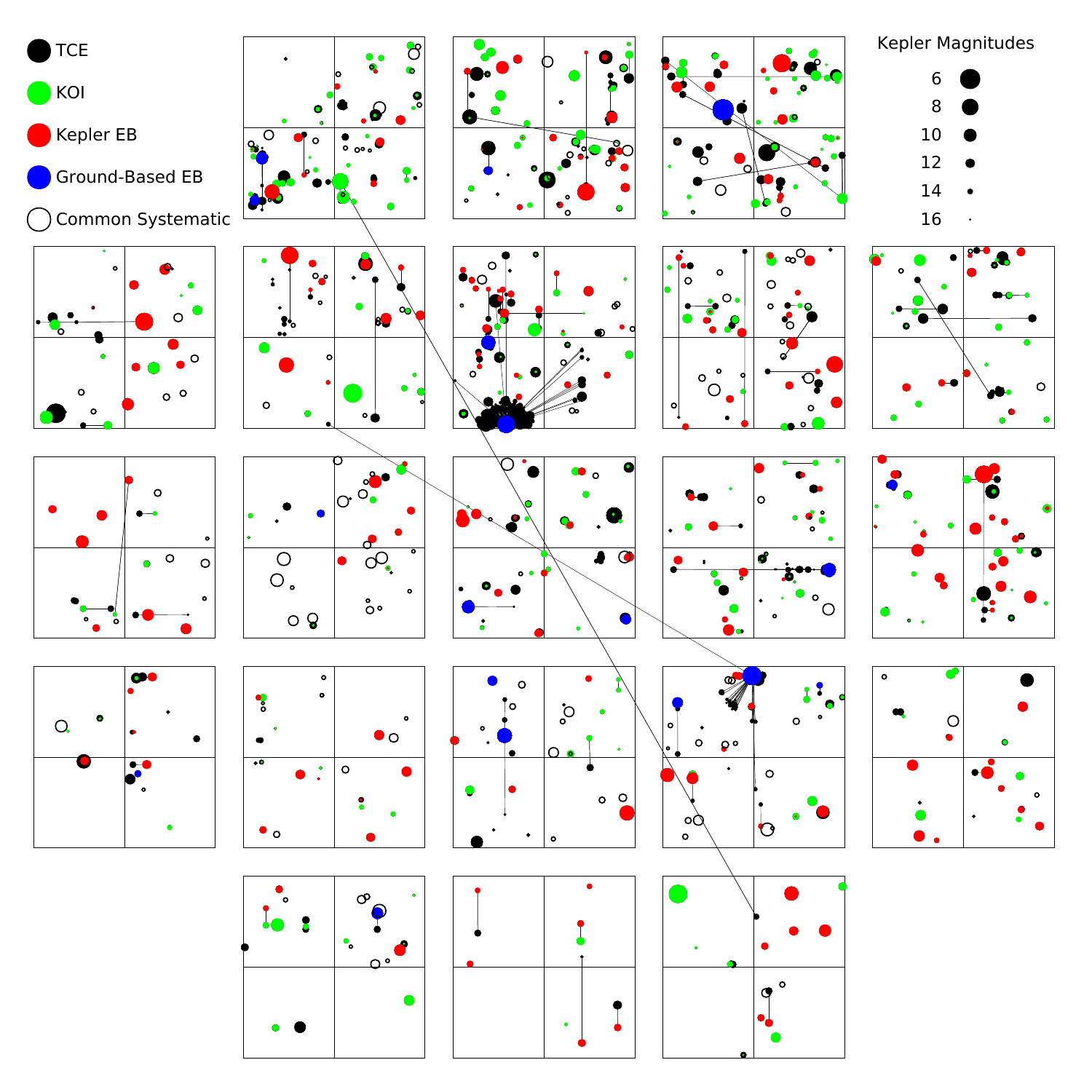}
\caption{Distribution of ephemeris matches on the focal plane. Symbol size scales with magnitude, while color represents the catalog in which the contaminating source was found. Blue indicates that the true transit is from a variable star only known as a result of ground-based observations. Red circles are stars listed in the \kepler{} EBWG catalog, green are KOIs, and black are TCEs. Open black points represent TCEs due to a common systematic. Black lines connect false positive matches with the most likely contaminating parent. In most cases parent and child are so close that the connecting line is invisible. Note that FP TCEs due to the common systematic are not connected by lines as they are not due to contamination from a variable source.}
\label{ephemmatchfig}
\end{figure*}

Via ephemeris matching, we identify \nephemmatch{} Q1--Q17~DR24 TCEs as FPs. Of these, \nonlyephemmatch{} were identified as FPs only due to ephemeris matching. We list all \nephemmatch{} TCEs in Table~\ref{ephemmatchtab}, as this information is valuable for studying contamination in the \kepler{} field. (Note that each TCE identified consists of its KIC ID and planet number, separated by a dash.) We also list in Table~\ref{ephemmatchtab} each TCE's most likely parent, the period ratio between child and parent (P$_{\rm rat}$), the distance between the child and parent in arcseconds, the offset in row and column between the child and parent in pixels ($\Delta$Row and $\Delta$Col), the magnitude of the parent (m$_{\rm Kep}$), the difference in magnitude between the child and parent ($\Delta$Mag), the depth ratio of the child and parent (D$_{\rm rat}$), the mechanism of contamination, and a flag to designate unique situations. In Figure~\ref{ephemmatchfig} we plot the location of each false positive TCE and its most likely parent, connected by a solid line. TCEs are represented by solid black points, KOIs are represented by solid green points, EBs found by \kepler{} are represented by solid red points, EBs discovered from the ground are represented by solid blue points, and TCEs due to a common systematic are represented by open black points. The \kepler{} magnitude of each star is shown via a scaled point size. Note that most parent-child pairs are so close together that the line connecting them is not easily visible on the scale of the plot.

\begin{deluxetable*}{ccccccccccc}
\tablecolumns{11}
\tabletypesize{\scriptsize}
\tablewidth{\linewidth}
\tablecaption{The \nephemmatch{} Q1--Q17~DR24 TCEs Identified as FPs due to Ephemeris Matches}
\tablehead{\colhead{TCE} & \colhead{Parent} & \colhead{P$_{\rm rat}$} & \colhead{Distance} & \colhead{$\Delta$Row} & \colhead{$\Delta$Col} & \colhead{m$_{\rm Kep}$} & \colhead{$\Delta$Mag} & \colhead{D$_{\rm rat}$} & \colhead{Mechanism} & \colhead{Flag} \\ & & & (\arcsec) & (Pixels) & (Pixels) & & & & & }
001295289-01 & \nodata & \nodata & \nodata & \nodata & \nodata & \nodata & \nodata & \nodata & Systematic & 0\\
002163326-01 & \nodata & \nodata & \nodata & \nodata & \nodata & \nodata & \nodata & \nodata & Systematic & 0\\
002166206-01 & 3735.01 & 1:1 & 8.3 & -1 & -2 & 17.64 & -4.34 & 3.4523E+02 & Direct-PRF & 0\\
002297793-01 & \nodata & \nodata & \nodata & \nodata & \nodata & \nodata & \nodata & \nodata & Systematic & 0\\
002305311-01 & 002305372-pri & 1:1 & 42.6 & 2 & 10 & 13.82 & 1.14 & 6.9390E+03 & Direct-PRF & 0\\
002308603-01 & \nodata & \nodata & \nodata & \nodata & \nodata & \nodata & \nodata & \nodata & Systematic & 0\\
002309585-01 & 5982.01 & 1:1 & 11.7 & -2 & 1 & 13.93 & 1.45 & 6.5525E+00 & Direct-PRF & 0\\
002437112-01 & 3598.01 & 1:1 & 19.7 & -5 & 1 & 17.63 & -1.48 & 7.0495E+02 & Direct-PRF & 0\\
002437112-02 & 3598.01 & 1:2 & 19.7 & -5 & 1 & 17.63 & -1.48 & 8.2520E+02 & Direct-PRF & 0\\
002437112-03 & 3712.01 & 1:1 & 15.1 & 4 & 1 & 16.99 & -0.84 & 7.6798E+02 & Direct-PRF & 0\\
\nodata & \nodata & \nodata & \nodata & \nodata & \nodata & \nodata & \nodata & \nodata & \nodata & \nodata
\enddata
\tablecomments{A suffix of ``pri'' in the parent name indicates the object is an EB known from the ground, and the child TCE matches to its primary. Similarly a suffix of ``sec'' indicates the child TCE matches the secondary of a ground-based EB. Parent names are listed, in priority order when available, by (1) their Bayer designation (e.g., RR-Lyr-pri), (2) their EBWG designation (e.g., 002305372-pri), (3) their KOI number (e.g., 3735.01), and (4) their TCE number (e.g., 002437452-01). A flag of 1 indicates that the TCE is a bastard, which are cases where two or more TCEs match each other via the Direct-PRF contamination mechanism, but neither can physically be the parent of the other via their magnitudes, depths, and distances, and thus the true parent has not been identified. A flag of 2 indicates cases of column anomalies that occur on different outputs of the same module. These cases likely involve cross-talk to carry the signal from one output to another. TCEs due to the common systematic do not have information listed for a parent source, as they are not caused by a single parent. Note that  Table~\ref{ephemmatchtab} is published in its entirety in the electronic edition of the Astrophysical Journal. A portion is shown here for guidance regarding its form and content.}
\label{ephemmatchtab}
\end{deluxetable*}

The larger number of matches compared to the Q1--Q12 and Q1--Q16 catalogs is predominately the result of a much larger short-period false positive population compared to Q1--Q16, and an extended baseline compared to Q1--Q12, coupled with matching all TCEs and not just KOIs. In Q1--Q17~DR24 we identify an additional contamination mechanism, which we label ``Common Systematic''. As mentioned in \S\ref{tcesec}, these are over 200 TCEs that are caused by 3 systematic events that are common to all \kepler{} CCDs and happen to be equidistant in time with a spacing of $\sim$459 days. 

We also identify 119 examples of ``Column Anomaly'', which is a previously identified mechanism where a parent is able to contaminate a child at large distances if they both lie on the same column of a CCD. This mechanism is particularly pernicious because it does not result in a visible centroid offset; the apparent location of the transit signal via the difference images coincides with the target. If the parent is not observed by \kepler{}, then the child could go undetected as a FP due to the column anomaly, as was recently the case for KOI~6705.01 \citep{Gaidos2015}. The large number of examples of column anomaly now available in the Q1--Q17~DR24 catalog reveals the following:

\begin{itemize}

\item Despite equally searching for matches in row and column, no instance of ``row-anomaly'' has been found to occur.

\item The CCDs are read out in the column direction.

\item In 91.6\% of cases, the child is at a higher row number than the parent, and thus the parent's pixels are read out before the child's. (The remaining 8.4\% of cases may not have the true parent identified, but rather a sibling, as only the most likely parent is listed, and many parents are unobserved by the spacecraft.)

\item Most cases show the depth of the child increases over time.

\item The effect appears to exhibit seasonal depth variations in most cases.

\item The average depth ratio between parent and child is a factor of $\sim$$10^{4}$, and typically the parent and child have similar magnitudes.

\end{itemize}

\noindent Combining these details leads to our conjecture that the column anomaly is due to decreasing charge transfer efficiency over time, likely due to cosmic ray impacts. When the CCD is read out, some charge from the parent is left behind due to charge transfer inefficiency. As the child is read out, and its electrons pass through the pixels where the parent was, the child picks up some of the parent's left-behind electrons. Thus, the variable signal from the parent is induced in the child. As more cosmic ray impacts accumulate over time, the amount of charge left behind by the parent increases, resulting in an increase in contamination, and thus an increase in the observed depth of the child. Seasonal variation is seen as the parent and child rotate between 4 CCDs with season, and the amount of degradation varies with CCD. The average depth ratio, along with the delta magnitudes observed, indicate that a charge transfer efficiency of $\sim$99.99\% is consistent with the observed contamination, i.e., a degradation of $\sim$0.01\%. This is well within the range observed on Hubble's Advanced Camera for Surveys and other spaceborne detectors  \citep[see \S3.7 of][and references therein]{Sirianni2005}.

\section{TCE Dispositioning and KOI Modeling}
\label{resultssec}

The robovetter was run on all \ntces{} Q1--Q17~DR24 TCEs. In the following subsections we describe the process of preparing the input for the robovetter, federating old and designating new KOI numbers, and modeling the KOIs to obtain planetary parameters with robust uncertainties.

\subsection{Robovetter Input}

In Table~\ref{roboinputtab} we list each of the \ntces{} Q1--Q17~DR24 TCEs and all the parameters that were used by the robovetter. These include:

\begin{itemize}
\item The period of the TCE in days, epoch in Barycentric \kepler{} Julian Date (BKJD), and duration of the transit in hours, all from the DV module of the \kepler{} pipeline.
\item The MES, and maximum SES value used in determining the MES, from the \kepler{} pipeline.
\item The LPP value using the DV detrending (LPP$_{\rm DV}$) and the alternate detrending (LPP$_{\rm Alt}$).
\item The value of the Marshall metric.
\item The odd/even depth statistic ($\sigma_{OE}$) for both the DV and alternate detrendings.
\item The metrics produced from the model-shift uniqueness test (see \S\ref{notuniquetcesec}) for both the DV and alternate detrending.
\item The radius of the planet in \re{}, calculated by multiplying the radius ratio from the DV module of the \kepler{} pipeline and the stellar radius value from \citet{Huber2014a}.
\item The albedo ($A$), primary depth ($D_{pri}$), secondary depth ($D_{sec}$), and secondary's phase ($Ph_{sec}$) for the DV and alternate detrendings (see \S\ref{secdetectsec}).
\item The disposition value from the centroid module. A value of 1 indicates a significant centroid offset was detected, while a value of 0 indicates no offset.
\end{itemize}

\begin{deluxetable*}{crccccccccc}
\tablecolumns{11}
\tabletypesize{\scriptsize}
\tablewidth{\linewidth}
\tablecaption{Robovetter Input Parameters for the \ntces{} Q1--Q17~DR24 TCEs}
\tablehead{\colhead{TCE} & \colhead{Period} & \colhead{Epoch} & \colhead{Duration} & \colhead{SES$_{\rm Max}$} & \colhead{MES} & LPP$_{\rm DV}$ & LPP$_{\rm Alt}$ & Marshall & $\sigma_{\rm OE,DV}$ & \colhead{\nodata} \\ & \colhead{(Days)} & \colhead{(BKJD)} & \colhead{(Hours)} & & & & & }
\startdata
000757450-01 & 8.884923 & 134.452041 & 2.078 & 1.130e+02 & 5.240e+02 & 2.370e-04 & 4.100e-05 & 0.000e+00 & 0.000e+00 & \nodata\\
000892667-01 & 2.262112 & 132.171131 & 7.509 & 3.890e+00 & 8.037e+00 & 4.608e-03 & 1.884e-03 & 0.000e+00 & 6.794e-02 & \nodata\\
000892772-01 & 5.092598 & 133.451376 & 3.399 & 3.810e+00 & 1.562e+01 & 1.337e-03 & 1.081e-03 & 0.000e+00 & 1.692e-01 & \nodata\\
001026032-01 & 8.460442 & 133.774329 & 4.804 & 4.957e+02 & 3.889e+03 & 4.660e-04 & 8.300e-05 & 0.000e+00 & 1.244e-01 & \nodata\\
001026032-02 & 4.230222 & 133.998093 & 4.606 & 1.704e+02 & 1.440e+03 & 3.030e-04 & 3.900e-05 & 0.000e+00 & 0.000e+00 & \nodata\\
001026133-01 & 1.346292 & 132.841605 & 1.626 & 4.530e+00 & 1.051e+01 & 3.540e-03 & 6.126e-03 & 0.000e+00 & 6.815e-02 & \nodata\\
001026133-02 & 2.691910 & 132.267127 & 5.530 & 3.320e+00 & 1.135e+01 & 4.856e-03 & 7.933e-03 & 0.000e+00 & 6.316e-02 & \nodata\\
001026957-01 & 21.761298 & 144.779125 & 1.277 & 2.383e+01 & 1.034e+02 & 2.570e-04 & 1.230e-04 & 0.000e+00 & 1.615e-01 & \nodata\\
001028018-01 & 0.614378 & 131.652061 & 1.448 & 6.850e+00 & 1.281e+01 & 6.614e-03 & 6.647e-03 & 0.000e+00 & 8.698e-04 & \nodata\\
001160891-01 & 0.940463 & 132.400156 & 3.354 & 4.010e+00 & 1.203e+01 & 7.627e-03 & 2.202e-03 & 0.000e+00 & 7.456e-03 & \nodata\\
\nodata & \nodata & \nodata & \nodata & \nodata & \nodata & \nodata & \nodata & \nodata & \nodata & \nodata\\
\enddata
\tablecomments{Table~\ref{roboinputtab} is published in its entirety in the electronic edition of the Astrophysical Journal. A portion is shown here for guidance regarding its form and content.}
\label{roboinputtab}
\end{deluxetable*}

\subsection{KOI Federation and New KOI Designation}
\label{koisec}

Transit-like signals found over the course of the \kepler{} mission are given \kepler{} Object of Interest (KOI) numbers in order to facilitate the tracking of these objects over multiple runs of the \kepler{} pipeline. Using the same procedure as \citet[][see \S4.1]{Mullally2015a}, which examines the number of overlapping in-transit cadences between two ephemerides, we federate \nfedkois{} Q1--Q17~DR24 TCEs to existing KOIs. 

Given that there were \npredrtwentyfourkois{} KOIs known prior to the Q1--Q17 DR 24 pipeline run, this indicates, at first glance, a 81.5\% KOI recoverability rate. Unrecovered KOIs can be planets in systems with large transit-timing variations, or transit-like systems in regions of parameter space that are affected by \kepler{} pipeline changes (see \S\ref{tcesec}). However, some unrecovered KOIs could have been dispositioned as not transit-like false positives after being promoted to KOI status in previous catalogs, and thus are rightfully not recovered by the pipeline due to additional data and improvements in the data processing and detection algorithms. (As a rule, once a KOI number is designated, it is never removed from the catalog.) Thus, given that there were \npredrtwentyfourtlkois{} transit-like KOIs known prior to the Q1--Q17 DR 24 pipeline run, of which \ntlfedkois{} federated, this indicates a 90.2\% transit-like KOI recoverability rate.

With respect to the Q1--Q17~DR24 robovetter, we assign new KOI numbers to nearly all transit-like TCEs (i.e., those that were \emph{not} designated not transit-like) that did not federate with previously known KOIs. New KOIs on stars that had previously associated KOIs were given the same base KOI number with the next-highest unused planet number. New KOIs on stars that did not have any previously associated KOIs were given numbers 6252.01 and higher. The only TCEs deemed transit-like by the robovetter that did not receive KOI numbers were \nbankois{} systems, listed in Table~\ref{bankoistab}. These systems were so complicated or unusual (e.g., extreme TTV systems, circumbinary planets, seasonally dependent contamination, severe detrending issues) that the resulting TCEs did not accurately correspond to the underlying transit-like signal. In total, we created \nnewkois{} new KOIs, thus extending the total number of KOIs from all KOI catalogs to \ntotkois{}. Note that while developing the robovetter, some KOI numbers were assigned to TCEs that were initially dispositioned as transit-like, but due to code changes, were later dispositioned as not transit-like, and thus there are some new KOIs in this catalog that are dispositioned as not transit-like FPs.

\begin{deluxetable}{c}
\tablecolumns{1}
\tablewidth{\linewidth}
\tablecaption{The \nbankois{} Anomalous TCEs that were Deemed Transit-Like by the Robovetter but were not Assigned KOI Numbers}
\tablehead{\colhead{Q1--Q17~DR24 TCE}}
002157247-01\\
003098184-01\\
003650049-01\\
004247023-01\\
004384675-04\\
005983532-01\\
006462874-01\\
006462874-02\\
006762829-03\\
006762829-04\\
006964043-01\\
007024511-01\\
007918172-01\\
007918172-02\\
008009496-01\\
008414907-01\\
008435232-01\\
009032900-02\\
009902856-01\\
009957659-01\\
010223616-01\\
010743597-04\\
011513441-01\\
012644769-03\\
012644774-01
\enddata
\label{bankoistab}
\end{deluxetable}

In Table~\ref{robodispstab} we list all \ntces{} TCEs, their assigned KOI numbers (if transit-like), their robovetter dispositions (PC or FP), the values of the four major flags (as described in \S\ref{robosec}), and a comment field that has mnemonic flags that describe the result of each individual robovetter test. Detailed descriptions for each mnemonic flag are located in Appendix~\ref{minorflagsec}. Note that every FP will have at least one major flag set, and could have any combination of all four. When both the not transit-like and significant secondary flags are set, it indicates that the TCE corresponds to the secondary eclipse of a system (e.g., TCE 001026032-02 in Table~\ref{robodispstab}). While we do not assign new KOI numbers to TCEs that are dispositioned as secondary eclipses in this catalog, there are pre-existing KOIs that federate with Q1--Q17~DR24 TCEs dispositioned as secondary eclipses. PCs will not have any major flags set, unless the system is a hot Jupiter with a visible secondary eclipse due to planetary reflection and/or thermal emission, in which case the significant secondary flag will be set. This information is also publicly available at the NASA Exoplanet Archive in the Q1--Q17~DR24 KOI table.

\begin{deluxetable*}{cccccccl}
\tablecolumns{8}
\tabletypesize{\scriptsize}
\tablewidth{\linewidth}
\tablecaption{Robovetter Dispositions, Major Flags, and KOI Numbers for the \ntces{} Q1--Q17~DR24 TCEs}
\tablehead{\colhead{TCE} & \colhead{KOI} & \colhead{Disp} & \colhead{N} & \colhead{S} & \colhead{C} & \colhead{E} & \colhead{Comments}}
\startdata
000757450-01 & 0889.01 & PC & 0 & 0 & 0 & 0 & \nodata\\
000892667-01 & \nodata & FP & 1 & 0 & 0 & 0 & LPP\_DV\_TOO\_HIGH\\
000892772-01 & 1009.01 & FP & 0 & 0 & 1 & 0 & CLEAR\_APO\\
001026032-01 & 6252.01 & FP & 0 & 1 & 0 & 0 & SIG\_SEC\_IN\_DV\_MODEL\_SHIFT---SIG\_SEC\_IN\_ALT...\\
001026032-02 & \nodata & FP & 1 & 1 & 0 & 0 & THIS\_TCE\_IS\_A\_SEC\\
001026133-01 & \nodata & FP & 1 & 0 & 0 & 0 & LPP\_DV\_TOO\_HIGH---LPP\_ALT\_TOO\_HIGH---ALT\_SI...\\
001026133-02 & \nodata & FP & 1 & 0 & 0 & 0 & LPP\_DV\_TOO\_HIGH---LPP\_ALT\_TOO\_HIGH---ALT\_SI...\\
001026957-01 & 0958.01 & PC & 0 & 0 & 0 & 0 & KIC\_OFFSET\\
001028018-01 & \nodata & FP & 1 & 0 & 0 & 0 & LPP\_DV\_TOO\_HIGH---LPP\_ALT\_TOO\_HIGH---EYEBALL...\\
001160891-01 & \nodata & FP & 1 & 0 & 0 & 0 & LPP\_DV\_TOO\_HIGH---DV\_SIG\_PRI\_OVER\_FRED\_TOO...\\
\nodata & \nodata & \nodata & \nodata & \nodata & \nodata & \nodata & \nodata
\enddata
\tablecomments{For the four major flags, Not Transit-Like is abbreviated as ``N'', Significant Secondary is abbreviated as ``S'', Centroid Offset is abbreviated as ``C'', and Ephemeris Match is abbreviated as ``E''. The mnemonic flags in the comments column are separated by dashes, and described in Appendix~\ref{minorflagsec}. Table~\ref{robodispstab} is published in its entirety in the electronic edition of the Astrophysical Journal. A portion is shown here for guidance regarding its form and content.}
\label{robodispstab}
\end{deluxetable*}

\subsection{KOI Modeling}
\label{koimodelsec}

In order to obtain transit model fits with robust uncertainties, we model every KOI in the same manner as described in \citet[][see \S5]{Rowe2015a} and \citet[][see \S6.2]{Mullally2015a}. To summarize briefly, we fit all available PDC data from DR24 at MAST after detrending via a polynomial filter as described in \S4 of \citet{Rowe2014}. We use the transit model of \citet{Seager2003}, assuming a circular orbit, with the quadratic limb-darkening law of \citet{Claret2011}, calculated for the \kepler{} passband. Uncertainties are calculated using a Markov Chain Monte Carlo \citep[MCMC;][]{Ford2005} method with four chains of $10^{5}$ fits each, discarding the first 20\% of each chain, to construct the posterior distributions. The transit model fit parameters are then combined with the stellar parameters to produce planetary parameters. The MCMC chains are publicly available and documented in \citet{Rowe2015b}.

KOIs which existed prior to Q1--Q17~DR24 were not re-fit in this work, and thus use stellar values from the Q1--Q16 stellar catalog \citep{Huber2014a} and contain values for their fit parameters identical to those given in \citet{Mullally2015a}. Newly designated KOIs are fit using the DR24 light curves and use stellar values from the updated Q1--Q17~DR24 stellar catalog \citep{Huber2014b}. The best-fit value and 1$\sigma$ uncertainties of each parameter are listed at the NASA Exoplanet Archive, along with the MCMC chains themselves. Note that not all KOIs were able to be modeled, which typically occurs when the polynomial filter (a separate detrending used specifically for the MCMC fitting) does not recover the transit events with sufficient SNR. These cases are designated in the KOI catalog by a value of ``none'' for the ``fittype'' parameter, and only the period, epoch, and duration of the federated TCE is reported.

\section{Analysis of the Q1--Q17~DR24 Catalog}
\label{analsec}

In order to be confident that the robovetter is properly reproducing the results of human TCERT members, it is informative to compare the Q1--Q17~DR24 KOI catalog to past KOI catalogs. Also, there are several ancillary \kepler{} catalogs that provide valuable checks on the quality of the KOI catalog. The injection of artificial transits into the \kepler{} pixel-level data also provides a valuable diagnostic of the performance of the robovetter and the completeness of the KOI catalog. Examining the results with respect to single- and multi-planet systems is yet another check to ensure the fidelity of the catalog. Finally, detecting potentially rocky planets that are possibly in the habitable zone of their host star is \keplers{} primary science goal, and as such those candidates are given extra scrutiny.

\subsection{Comparison to Past KOI Catalogs}

Of the \ntlfedkois{} transit-like KOIs that existed prior to the Q1--Q17~DR24 activity and were detected as TCEs by the Q1--Q17~DR24 \kepler{} pipeline, \ntlfedkoisrvtl{} were dispositioned by the robovetter as transit-like, yielding a 97.4\% transit-like KOI recoverability rate for the robovetter. Similarly, of the \npredrtwentyfourfedpcs{} pre-existing PCs that were re-detected, \npredrtwentyfourfedpcsrvpc{} were dispositioned as PCs, yielding a 96.9\% PC robovetter recoverability rate. Finally, of the \npredrtwentyfourfedfps{} pre-existing FPs that were re-detected, \npredrtwentyfourfedfpsrvfp{} were dispositioned by the robovetter as FPs, yielding a 89.3\% FP robovetter recoverability rate. 

Compared to past catalogs, the dispositions of \npctofp{} KOIs changed from PC to FP, and \nfptopc{} went from FP to PC. Examining these KOIs, we note that many changed dispositions due to the robovetter out-performing the human vetters. For example, the robovetter reliably detects very small secondary eclipses that the humans tended to miss. Also, the robovetter does not declare FPs based on transit depth alone, which was a directive given to the human vetters, but not followed by all vetters. Thus, the Q1--Q17~DR24 catalog contains more PCs with very deep depths compared to previous catalogs. We also note that the Q1--Q6 and Q1--Q8 catalogs were not solely based on the TCE list from the \kepler{} pipeline, and included KOIs found by other transit search techniques as well as manual light curve inspection.

\subsection{Comparison to Ancillary \kepler{} Catalogs}
\label{ancilsec}

\subsubsection{The Eclipsing Binary Working Group Catalog}

The EBWG catalog is the result of years of effort by the EBWG to identify and classify every eclipsing binary observed by \kepler{} \citep{Prsa2011,Slawson2011,Kirk2015}, and provides a valuable test of the efficiency of the robovetter in detecting EBs. We searched the EBWG catalog for systems with visible secondaries, since the robovetter is designed to only disposition EBs as FPs if a distinct, significant secondary event is detected. (FPs are purposely not designated based on depth or inferred size alone --- see \S\ref{sigsecsec}). At the time of closing the Q1--Q17~DR24 KOI table, there were 933 detached eclipsing binaries in the EBWG catalog with a distinct secondary eclipse, as defined by a EBWG morphology parameter less than 0.6 \citep{Matijevic2012} and a secondary eclipse that is either offset from phase 0.5 by at least 0.01 phase or has a depth at least 10\% different than the primary.

Of these 933, 894 are detected as TCEs by the Q1--Q17~DR24 \kepler{} pipeline, yielding a \kepler{} pipeline EB detection efficiency of 95.8\%. Examining the 39 that were not detected, they appear to have either (1) very low SNR, (2) very short periods and shapes such that the harmonic remover may have suppressed their signal, or (3) extremely long periods such that less than three primary transits are visible, as is required for a TCE detection. Of the 894 that were detected as TCEs, the robovetter designates 805 as FPs specifically due to a significant secondary, yielding a robovetter EB detection rate of 90.0\%. Of the 89 that the robovetter did not explicitly label EB, 40 were labeled not transit-like FPs, principally by the LPP metric, thus still yielding a robovetter FP detection rate of 94.5\%. The remaining 49 systems were principally called planet candidates due to either (1) detrending that significantly suppressed the depth of the secondary or (2) detection by the \kepler{} pipeline at half the orbital period, with the resulting odd/even difference not detected by the robovetter.

We also note that the EBWG often draws upon the results of the TCERT vetting from each catalog, and after performing their own vetting procedure, may incorporate them into the EBWG catalog. Prior to closing the Q1--Q17~DR24 KOI table, we found that the robovetter had identified several hundred TCEs as on-target EBs that were not yet cataloged by the EBWG. The list of these potentially new EBs was sent to the EBWG who then incorporated many of them into the EBWG catalog, prior to performing the comparison above.

\subsubsection{The False Positive Working Group Catalog}

The False Positive Working Group (FPWG) is manually vetting every KOI previously identified as a FP, along with a subset of PCs, to create the FPWG catalog \citep{Bryson2015}. Unlike TCERT, the FPWG takes a best-knowledge approach, using any and all available pieces of information to vet each KOI, including follow-up observations. This also includes designating false positives on the basis of transit depth, or inferred planetary radius, alone. We select the 1,346 certified FPs from the FPWG table, at the time of closing the Q1--Q17~DR24 KOI table, which federate to Q1--Q17~DR24 TCEs and have inferred planetary radii less than 25~\re{}. Of the 1,346, the robovetter designates 1,253 as FPs, yielding a 93.1\% rate of agreement.

\subsubsection{The \kepler{} Autovetter}

Another ancillary catalog generated for the Q1--Q17~DR24 activity is the ``autovetter'' catalog \citep{McCauliff2015,Catanzarite2015}, which uses a random forest machine learning approach to automatically classify TCEs based on training sets from previous KOI catalogs, using metrics from both DV and TCERT. It classifies each TCE into one of three categories: planet candidate (PC), astrophysical false positive (AFP), or non-transiting phenomenon (NTP). It defines PCs as TCEs that are consistent with a transiting planet, AFPs as TCEs that are due to detached or contact eclipsing binaries, pulsating stars, starspots, and other periodic signals of astrophysical origin, and NTPs as TCEs that are of instrumental or systematic origin.

There are 3,900 Q1--Q17~DR24 TCEs that the autovetter labels PC, of which the robovetter designates 3,775 as PC, for an agreement rate of 96.8\%. There are 16,467 TCEs that the autovetter labels AFP or NTP, of which the robovetter designates 15,944 as FP, for an agreement rate of 96.8\%. However, it is difficult to compare the AFP and NTP categories to any of the four major robovetter false positive flags, as the robovetter considers contact eclipsing binaries, pulsating stars, starspots, and other quasi-sinusoidal signals, along with instrumental noise, to be not transit-like, while the autovetter only considers instrumental noise to be non-transiting phenomenon.

\subsubsection{Planet Hunters}

As part of the Zooniverse citizen science platform \citep{Simpson2014}, Planet Hunters (PH) is a project where humans visually check \kepler{} light curves to search for transit signals, especially those not detected by the \kepler{} pipeline. We compiled a list of 63 planet candidates published by Planet Hunters \citep{Fischer2012,Lintott2013,Schwamb2013,Wang2013,Schmitt2014a,Schmitt2014b} and compare them to the Q1--Q17~DR24 KOI Catalog. Of the 63, 38 were detected as TCEs at the period identified by PH and were dispositioned as PC, 4 were detected as TCEs at the period identified by PH, but dispositioned as FP, 8 had TCEs detected around the same target, but not at the period identified by PH, and 13 had no TCEs detected around the target. Of the 4 that were identified at the same period, but were declared false positives by the robovetter, one was deemed not transit-like due to the Marshall metric, one was deemed to have a secondary eclipse by the model-shift test on the DV detrending, and the other two were deemed to have centroid offsets. The remaining PH candidates appear to mostly be planets around binary stars and in multi-planet systems with strong TTVs, or have very long periods such that three transits may not be visible, and thus are not expected to be detected as TCEs by the \kepler{} pipeline. However, these systems are extremely interesting scientifically, so the PH work highlights the importance of manual inspection in a dataset as rich and complex as that from \kepler{}.

\subsubsection{Confirmed Planets}
\label{confirmsec}

The NASA Exoplanet Archive designates some KOIs as ``confirmed planets'' based on the results of follow-up observations published in the literature. The follow-up observations may directly determine the mass of the planet via radial-velocity measurements, statistically validate the planet by fully characterizing the host star and any possible nearby sources of contamination, or in any other way demonstrate evidence for a planetary origin of the transit signal at the $\sim$99\% confidence level. The designation of confirmed planets by the Exoplanet Archive is completely independent of the PC/FP disposition given by TCERT, which is based solely on \kepler{} data.

Of the \nconfirmed{} confirmed \kepler{} planets that were listed at the NASA Exoplanet Archive, at the time of closing the Q1--Q17~DR24 KOI table, and that federate with Q1--Q17~DR24 TCEs, the robovetter designates \nconfirmedpcs{}, or 99.1\%, as PCs.  Of the nine confirmed planets that were designated FPs, two were dispositioned not transit-like, four were dispositioned as having significant secondaries, and three were dispositioned as having a centroid offset. For the two FPs due to being not transit-like, one failed due to the LPP test, and the other due to the model-shift uniqueness test, in both cases using the DV detrending. Upon manual inspection, these transit signals seem to be distorted in the DV detrending so that they no longer appear transit-like, probably because of DV's harmonic remover. For the four FPs due to significant secondaries, two appear to be caused by poor detrending that mimicked the appearance of a secondary, and two are due to remaining systematics from strong TTVs. For the three FPs due to centroid offsets, two appear due to systematics resulting from strong TTVs in multi-planet systems, and the other one is due to very large proper motion of the target, which is a late-type M dwarf. 

In all nine cases, we conclude that these confirmed planets should have been dispositioned as PC. While we will strive to further improve the robovetter to disposition these confirmed systems correctly, overall these nine systems are outliers with no consistent cause, and the robovetter is very efficiently ($>$99\%) dispositioning confirmed planets as planet candidates.

\subsection{Artificial Transit Injection}
\label{injectsec}

The primary way of measuring the efficiency of a transit detection pipeline is to inject artificial transit signals into the calibrated pixel-level data, with a range of parameters, and determine what fraction are detected as a function of those parameters. For the \kepler{} pipeline, \citet{Christiansen2013b} measured the detection efficiency of individual transit events, finding they were generally recovered to a high fidelity of $\sim$99.7\%. \citet{Christiansen2015b} then extended this work by injecting full time-series transit signals into a year of \kepler{} data, and were able to map out the actual \kepler{} pipeline recoverability rate as a function of MES, which is crucial to accurately determining planet occurrence rates \citep{Burke2015}.

In Q1--Q17~DR24, artificial transits were injected into the entire \kepler{} dataset at the pixel level, one injected transiting planet signal per star, with periods between 0.5--500 days and planetary radii between 0.25--20~\re{} \citep{Christiansen2015a}. A small fraction of these were purposely injected up to $\sim$10\arcsec{} away from the target star, in order to simulate FPs due to a centroid offset. The exact same version of the \kepler{} pipeline that produced the Q1--Q17~DR24 TCEs was used to search this injected dataset. In total, there were \ninjecttces{} injected signals detected by the \kepler{} pipeline, which we refer to as ``injTCEs''.

\begin{figure*}
\centering
\begin{tabular}{cc}
   \includegraphics[width=0.5\linewidth]{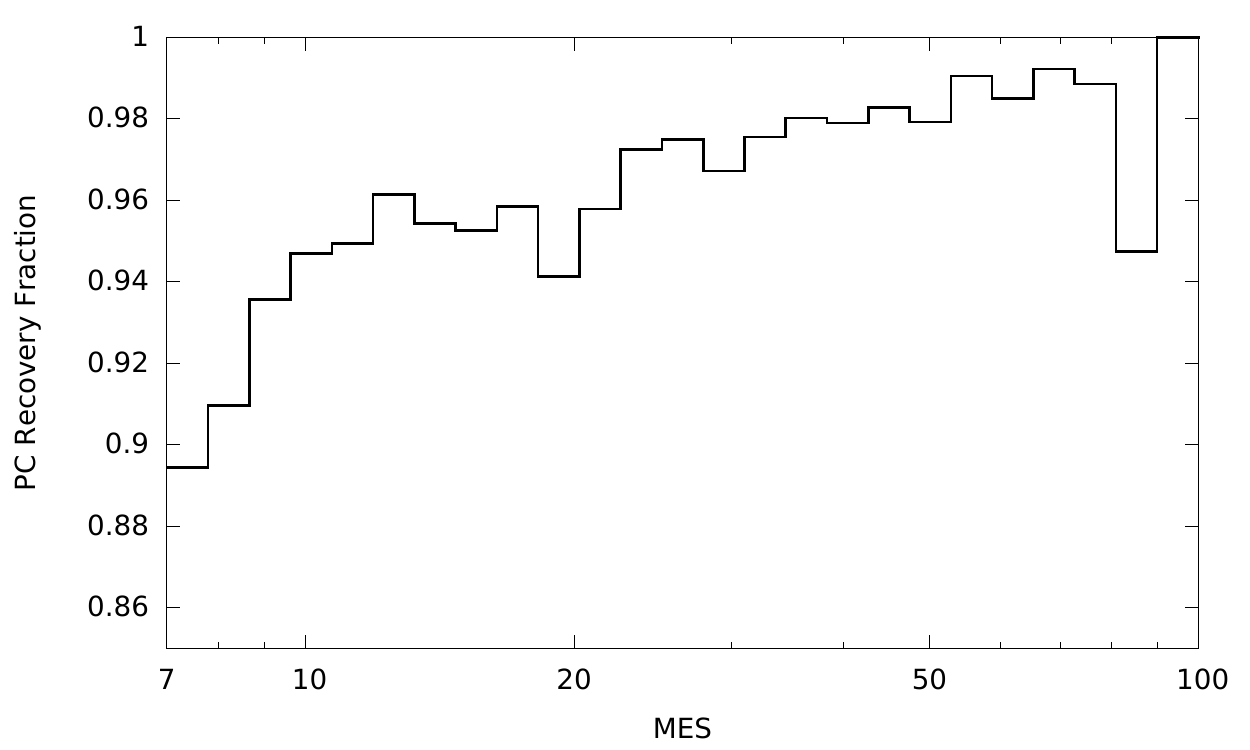} &
   \includegraphics[width=0.5\linewidth]{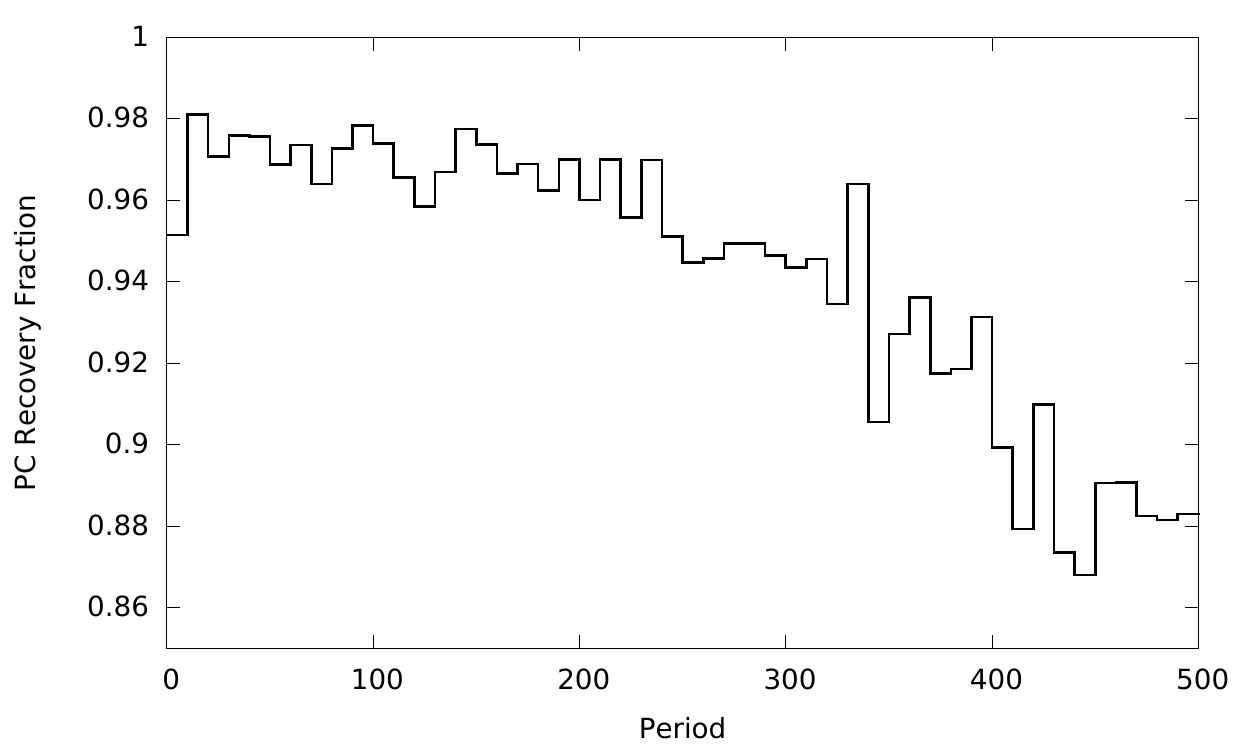} \\
   \includegraphics[width=0.5\linewidth]{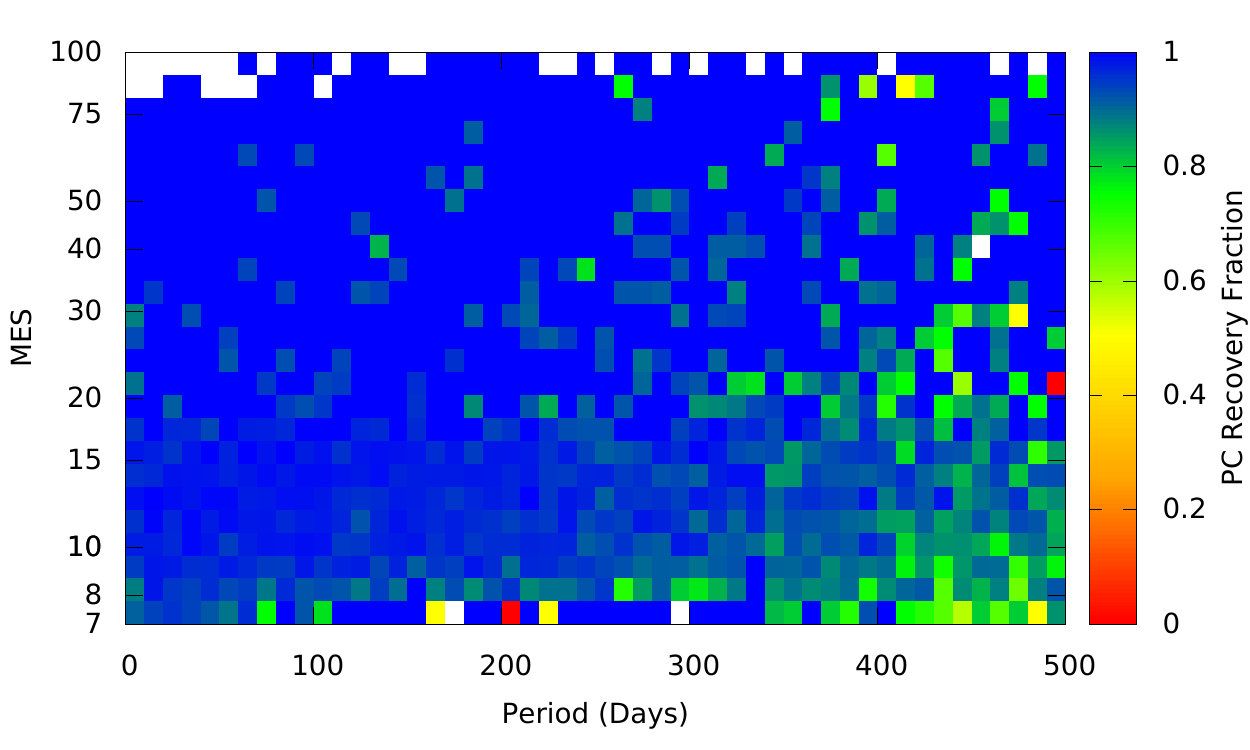} &
   \includegraphics[width=0.5\linewidth]{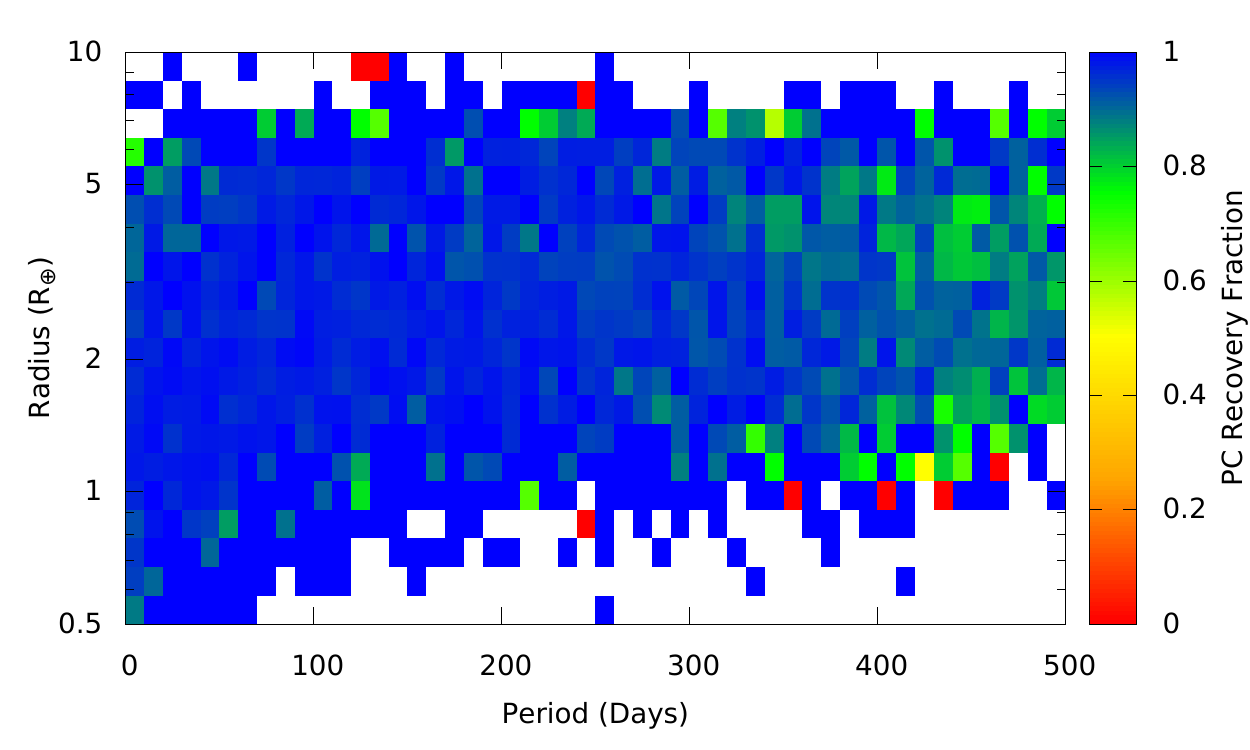} \\
   \includegraphics[width=0.5\linewidth]{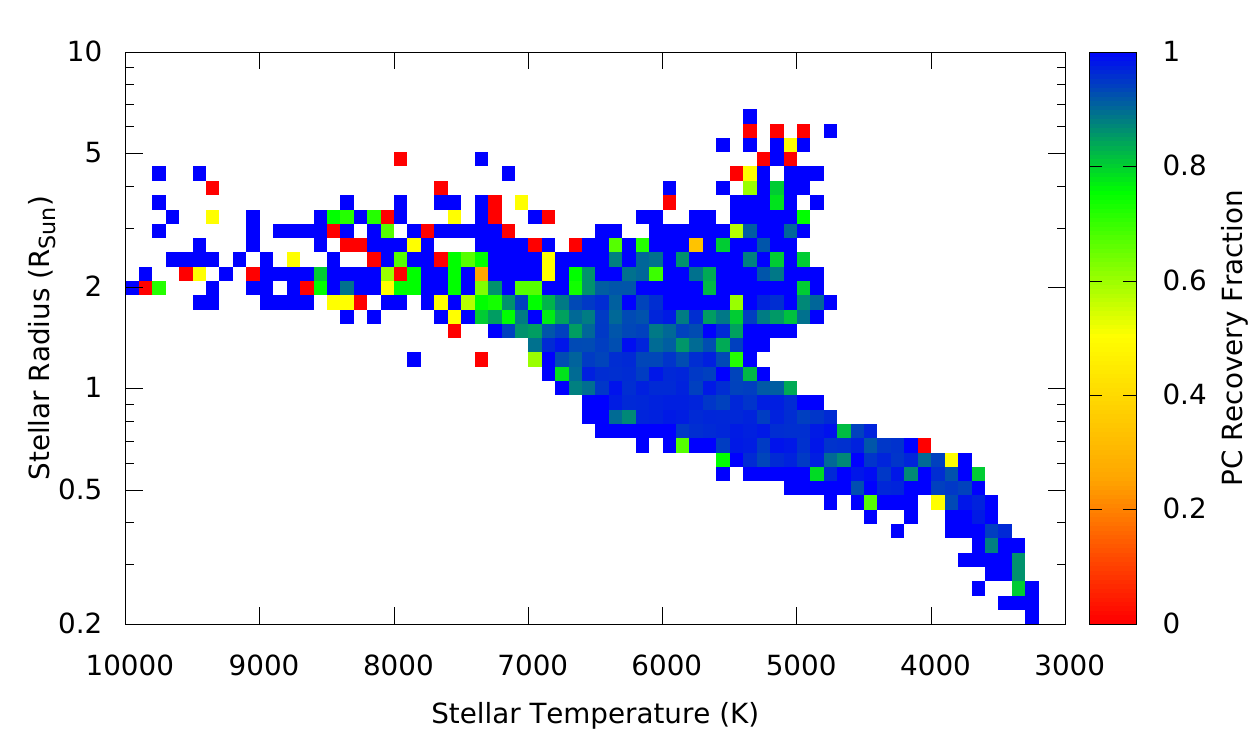} &
   \includegraphics[width=0.5\linewidth]{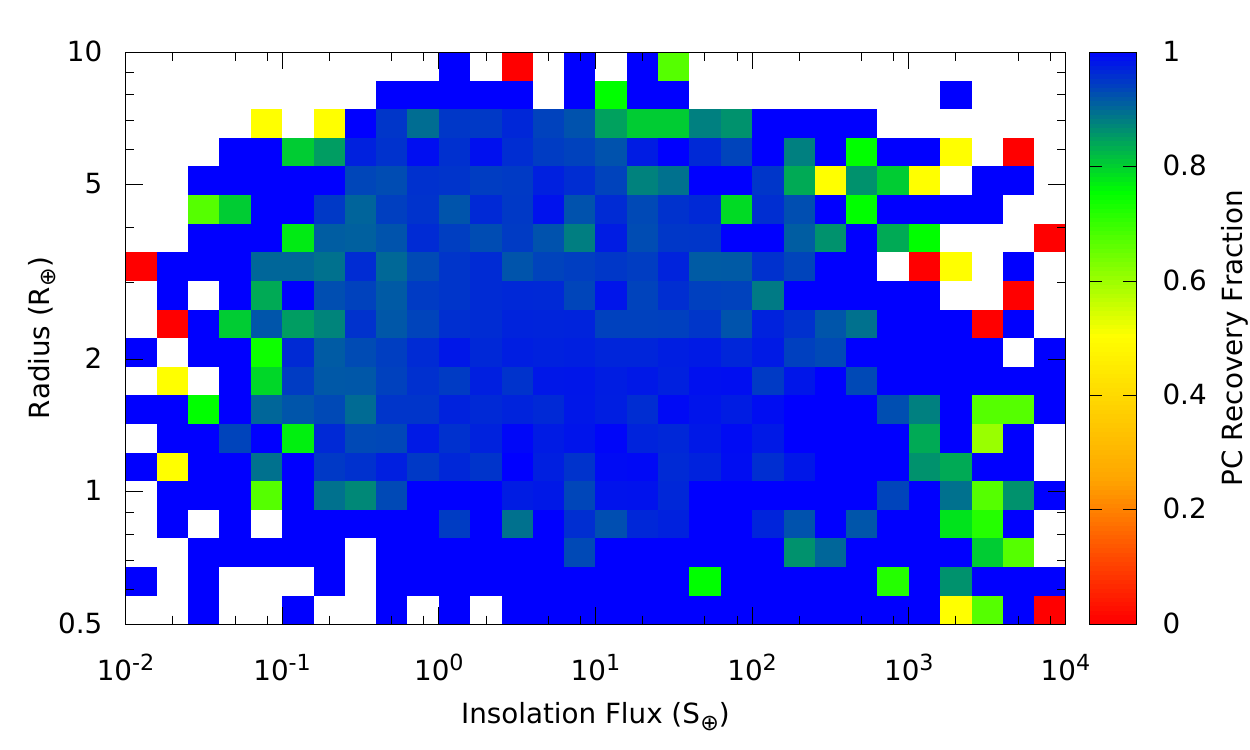}
\end{tabular}
\caption{The fraction of injected transit signals, recovered by the \kepler{} pipeline (i.e., injTCEs), that were labeled as PC by the robovetter. White areas represent bins where no injTCEs were detected. Top-left: The PC recovery fraction as a function of MES. Top-right: The PC recovery fraction as a function of period. Middle-left: The PC recovery fraction as a function of period and MES. Middle-right: The PC recovery fraction as a function of period and planet radius. Bottom-left: The PC recovery fraction as a function of host star radius and temperature. Bottom-right: The PC recovery fraction as a function of planet radius and insolation flux. Note that the insolation flux was calculated via $S$ = (\teq{}/255)$^{4}$, where $S$ is the insolation flux relative to the Earth, \teq{} is the equilibrium temperature of the planet in Kelvin as calculated by the \kepler{} pipeline, and 255~K is the Earth's equilibrium temperature.}
\label{tcert-detection-prob-fig}
\end{figure*}

We disposition the injTCEs with the exact same version of the robovetter that was used to disposition the Q1--Q17~DR24 TCEs \citep{Coughlin2015b}.  In Table~\ref{injecttab} we list each injTCE via its KIC number, along with the resulting robovetter disposition, major flags, and all injected and recovered parameters. In Figure~\ref{tcert-detection-prob-fig} we plot the fraction of on-target injTCEs that were labeled as PC by the robovetter (the PC recovery fraction) as functions of their MES, period, planetary radius, planetary insolation flux, stellar radius, and stellar temperature.

\begin{deluxetable*}{cccccccrcrrc}
\tablecolumns{12}
\tablewidth{\linewidth}
\tablecaption{Injected TCEs, Robovetter Dispositions, and Significant Parameters}
\tablehead{\colhead{KIC} & \colhead{Disp} & \colhead{N} & \colhead{S} & \colhead{C} & \colhead{E} & \colhead{Skygroup} & \colhead{injPeriod} & \colhead{injEpoch} & \colhead{injDepth} & \colhead{injDuration} & \colhead{\nodata} \\ & & & & & & & (Days) & (BKJD) & \colhead{(ppm)} & \colhead{(Hours)} & \colhead{\nodata}}
\startdata
1701692 &      PC &       0 &       0 &       0 &       0 &      71 & 357.1302 & 54933.9512 &    1224 &    7.31 & \nodata\\
1719026 &      PC &       0 &       0 &       0 &       0 &      84 &  92.2382 & 54912.5810 &     120 &    7.79 & \nodata\\
1719262 &      PC &       0 &       0 &       0 &       0 &      84 & 267.0962 & 55037.5158 &     164 &   10.29 & \nodata\\
1719371 &      PC &       0 &       0 &       0 &       0 &      84 &  96.4779 & 54934.2982 &    1324 &    3.43 & \nodata\\
1719472 &      PC &       0 &       0 &       0 &       0 &      84 & 287.7573 & 55174.7539 &     533 &   12.04 & \nodata\\
1719550 &      FP &       0 &       1 &       0 &       0 &      84 &  80.8994 & 54908.5248 &     238 &    5.75 & \nodata\\
1719927 &      PC &       0 &       0 &       0 &       0 &      84 & 282.5501 & 54980.8943 &    5009 &   10.58 & \nodata\\
1720670 &      PC &       0 &       0 &       0 &       0 &      84 &  84.3903 & 54926.2464 &    1117 &    5.24 & \nodata\\
1721110 &      PC &       0 &       0 &       0 &       0 &      84 &   3.2179 & 54900.7105 &      70 &    5.02 & \nodata\\
1721133 &      PC &       0 &       0 &       0 &       0 &      84 & 132.4065 & 54911.5644 &    1779 &    8.53 & \nodata\\
\nodata & \nodata & \nodata & \nodata & \nodata & \nodata & \nodata &  \nodata &    \nodata & \nodata & \nodata &   \nodata

\enddata
\tablecomments{Table~\ref{injecttab} is published in its entirety in the electronic edition of the Astrophysical Journal. A portion is shown here for guidance regarding its form and content.}
\label{injecttab}
\end{deluxetable*}

The robovetter dispositioned 34,210 of the \nnooffset{} $\rm injTCEs$ without centroid offsets as PC, yielding a 95.25\% pass rate. Examining Figure~\ref{tcert-detection-prob-fig}, specifically the top-left panel, it can be seen that the PC recovery fraction increases with increasing MES. (Note that the \kepler{} pipeline has a minimum detection threshold of 7.1 MES, and very few transit signals were injected with MES greater than 100.) While very low MES detections pass $\sim$90\% of the time, the highest MES detections pass $\sim$98\% of the time, as the vetting metrics become more reliable at higher MES values. Examining the top-right panel, the PC recovery fraction increases with decreasing period. (Note that no signals were injected with periods greater than 500 days.) These two trends can also be seen in the middle-left panel, where the PC recovery fraction is shown as a function of both period and MES, and in the middle-right panel, where the PC recovery fraction is shown as a function of planet radius and period. The bottom-left panel indicates that planets around higher-temperature and more evolved stars, particularly the instability strip at $\sim$7,500~K, may also have decreased PC recovery fractions compared to cooler, main-sequence stars, likely due to increased systematic noise from stellar pulsation that are not fully corrected by either of the two detrendings employed by TCERT.

Finally, we examine the PC recovery fraction of on-target injTCEs with radius (\rp{}) and insolation flux (\sp{}) values within 25\% of that of Earth's values (0.75~$>$~\rp{}~$>$~1.25~\re{} and 0.75~$>$~\sp{}~$>$~1.25~\se{}). There are 118 on-target injTCEs that meet these \rp{} and \sp{} criteria, of which 116 are designated PCs by the robovetter, therefore yielding a 98.3\% PC recovery fraction. This can be seen graphically in the bottom-right panel of Figure~\ref{tcert-detection-prob-fig}, where the area around Earth's values (1.0~\re{}, 1.0~\se{}) shows a very high PC recovery fraction. If we add the additional constraint that the host star's effective temperature (\tstar{}) is within 500K of the Sun's, (5300~$<$~\tstar{}~$<$~6300 K), in addition to the previous radius and insolation flux constraints, then the TCERT detection efficiency is 96.1\%, as 49 of 51 on-target injTCEs that meet these criteria are designated as PCs.

Note that one could make a robovetter with a 100\% detection efficiency by simply passing every TCE as a PC --- however this would be a very poor robovetter, as it would not identify any false positives! We have specifically designed the robovetter to identify as many false positives as possible while still correctly identifying at least $\sim$95\% of true planetary signals. This means that correcting for the robovetter's detection efficiency will only affect derived occurrence rates at the $\sim$5\% level for the entire population, which is small compared to other systematic effects that affect the determination of planetary occurrence rates \citep[see Figure 10 of][]{Burke2015}. We note that specific regions of interest may have higher or lower detection efficiencies.

At present (i.e., for DR24) we do not have a complete measure of how many true, underlying false positives the robovetter dispositions as planet candidates. This injection only run included signals purposely injected off-target to simulate FPs due to centroid offsets, and found a $\sim$50\% detection rate at a separation of 2$\arcsec$ (0.5 pixels) when recovered with a MES of 20 \citep{Mullally2015c}. To assess other types of FPs we recommend (1) injecting eclipsing binary signals to simulate FPs due to significant secondaries, (2) inverting the light curve and performing a transit search to simulate the population of not transit-like false positives, operating under the general observation that most not transit-like false positives tend to be symmetrical, and (3) shuffling the \kepler{} data by season and performing a transit search to simulate long-period false positives. Such activities are likely vital to fully evaluate the false positive rate of the \kepler{} pipeline and the robovetter, and thus determine accurate occurrence rates, especially for those with radii and insolation fluxes comparable to the Earth.

\subsection{Systems With Multiple Planet Candidates}

In the Q1--Q17~DR24 catalog there are a total of \npckois{} PCs in \npcsystems{} systems. Of these systems, \npcmultisystems{} contain two or more PCs, with a total of \npcmultikois{} PCs in multi-PC systems. Compared to past catalogs, looking at systems that increased in PC count, we find that:

\begin{itemize}
\item 47 systems went from 1 $\rightarrow$ 2 PCs.
\item 1 system went from 1 $\rightarrow$ 3 PCs.
\item 9 systems went from 2 $\rightarrow$ 3 PCs.
\item 4 systems went from 3 $\rightarrow$ 4 PCs.
\item 1 system went from 4 $\rightarrow$ 5 PCs.
\end{itemize}

\noindent The system of five PCs, KOI~4032, appears to be a particularly interesting compact multi-planet system, as all 5 planet candidates have periods between 2.9--7.2 days, with inferred radii between 0.8--1.0~\re{}, around a solar type star (5575~K, 1.06~\rsun{}).

Of the \nkois{} KOIs in the Q1--Q17~DR24 catalog, \nsingkois{} are in single KOI systems, and \nmultikois{} are in multi-KOI systems (at least 2 KOIs associated with the same target.) Of the \nsingkois{} KOIs in single systems, \npcsingkois{} are dispositioned as PC and \nfpsingkois{} as FP, yielding a 51.6\% FP rate. Of the \nmultikois{} KOIs in multiple systems, \npcmultikois{} are dispositioned as PC and \nfpmultikois{} as FP, yielding a 8.6\% FP rate. The lower FP rate is expected for multi-KOI systems, as systems with multiple KOIs are more likely to contain actual PCs \citep{Rowe2014}. While expected, this analysis provides a valuable check that the robovetter is not dispositioning a significant number of KOIs as FPs simply due to the fact they are in multi-KOI systems.

\subsection{Potentially Rocky Planets in the Habitable Zone}
\label{hzsec}

\begin{figure*}
\centering
\includegraphics[width=\linewidth]{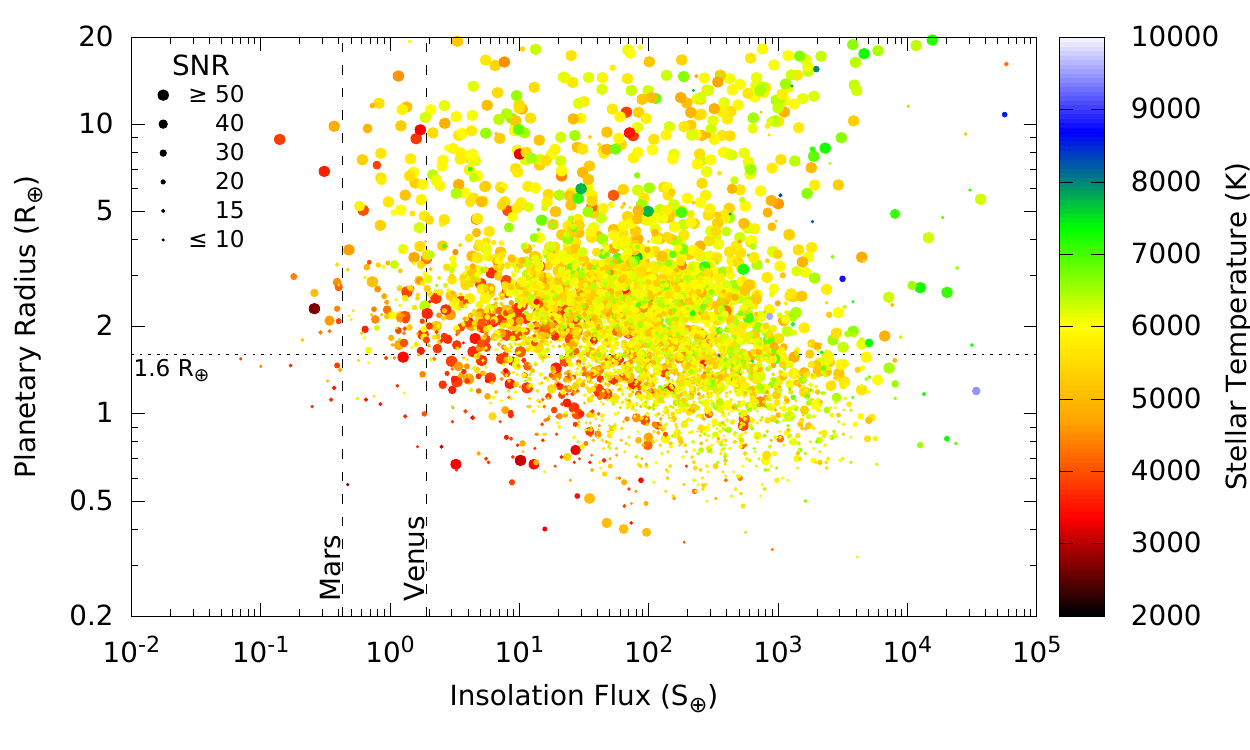}
\caption{A plot of planet radius versus insolation flux for all planet candidates known in the Q1--Q17~DR24 KOI catalog. (Note that some planet candidates, particularly those at large radii, lie outside the chosen axis limits for the plot, and thus are not shown.) The temperature of the host star is indicated via the color of each point, and the signal-to-noise of the detection is indicated via the size of each point. The two vertical dashed lines indicate the insolation flux values of Mars and Venus as a broad guide to a potential habitable zone. The horizontal dotted line is set at 1.6~\re{} as a suggested guide to where roughly half of the planets are expected to be rocky \citep{Rogers2015}.}
\label{rtss-fig}
\end{figure*}

In Figure~\ref{rtss-fig} we plot every Q1--Q17~DR24 TCE that was dispositioned as a PC by the robovetter as a function of its inferred planetary radius (\rp{}) and insolation flux (\sp{}). We also utilize point size to represent the SNR of each candidate, and the color of the point to indicate the effective temperature of the host star. We use vertical dashed lines to indicate the insolation flux levels of Mars and Venus, as a broad guide to a potential habitable zone. We use a horizontal dashed line to mark the radius at which a planet has about an even chance of being a terrestrial, rocky planet \citep{Rogers2015}.

As can be seen, while there are thousands of planet candidates, only a small percentage lie within the potential habitable zone. Smaller planets with lower insolation flux levels are predominately found around late-type stars. This is primarily an observational bias, as planets with shorter periods and larger radii relative to their host stars are more easily detected. Many of the small, low insolation flux planets have a SNR of $\sim$10 or less. In this low SNR regime, the odds of the TCE being a false positive that is undetectable by the robovetter is enhanced, and thus these candidates should continue to be treated with caution. More work is needed to obtain a quantitative measure of the rate of undetected low SNR false positives residing in the catalog.

Potentially rocky, habitable planets are the most important targets for follow-up observations to determine the frequency of Earth-size planets in the habitable zones of other stars. In Table~\ref{hzearthstab} we list all PCs in the Q1--Q17~DR24 catalog that have \rp{}~$<$~2.0~\re{} and \sp{}~$<$~2.0~\se{}. We list the values for their transit-model SNR, inferred planet radius, insolation flux, and host star effective temperature and radius from the Q1--Q17~DR24 KOI catalog. In addition, as discussed in \S\ref{confirmsec}, the NASA Exoplanet Archive maintains a list of KOIs that have been confirmed as planets via follow-up observations and/or statistical analyses, and assigns them \kepler{} confirmed planet numbers, e.g., Kepler-1b. (Again, note that the KOI PC/FP disposition and the NExScI confirmed planet designation are completely independent.) If the planet is listed as a confirmed planet at the NASA Exoplanet Archive confirmed planets table, we also list its \kepler{} confirmed planet number, reference for the confirmation, and values for the planetary radius, planetary insolation flux, and host star effective temperature and radius from the reference. If insolation flux was not given in the reference, we derive it from other values given in the reference, via,

\begin{equation}
S_{p} = \frac{R_{p}^{2} \cdot (T_{\star}/5777)^{4}}{a^{2}}
\end{equation}

\noindent where $a$ is the semi-major axis of the planet's orbit in AU, \tstar{} is in Kelvin, 5777~K is the effective temperature of the Sun, and \sp{} and \rp{} are in Earth units.

\begin{deluxetable*}{rrcccclccccr}
\tablecolumns{12}
\tablewidth{\linewidth}
\tablecaption{Small Planet Candidates Potentially in the Habitable Zone in the Q1--Q17~DR24 Catalog\\(\rp{}~$<$~2.0~\re{}~and \sp{}~$<$~2.0~\se{})}
\tablehead{ & & \multicolumn{4}{c}{Catalog Values} & & \multicolumn{4}{c}{Confirmed Values} & \\
\colhead{KOI} & \colhead{SNR} & \colhead{\rp{}} & \colhead{$S$} & \colhead{\tstar{}} & \colhead{\rstar{}} & \colhead{Confirmed} & \colhead{\rp{}} & \colhead{$S$} & \colhead{\tstar{}} & \colhead{\rstar{}} & \colhead{Reference}\\
& & \colhead{(\re{})} & \colhead{(\se{})} & \colhead{(K)} & \colhead{(\rsun{})} & \colhead{Name} & \colhead{(\re{})} & \colhead{(\se{})} & \colhead{(K)} & \colhead{(\rsun{})} & }
\startdata
172.02 & 20.7 & 1.74 & 1.59 & 5637 & 0.94 & Kepler-69c & 1.71 & 1.92 & 5638 & 0.93 & \citet{Barclay2013}\\
438.02 & 36.9 & 1.76 & 1.28 & 3985 & 0.54 & Kepler-155c & 2.24 & 2.43 & 4508 & 0.62 & \citet{Rowe2014}\\
463.01 & 72.1 & 1.57 & 1.26 & 3387 & 0.30 & \nodata & \nodata & \nodata & \nodata & \nodata & \nodata\\
571.05 & 12.4 & 1.06 & 0.25 & 3761 & 0.46 & Kepler-186f & 1.17 & 0.30 & 3755 & 0.52 & \citet{Torres2015}\\
701.03 & 45.0 & 1.73 & 1.17 & 4797 & 0.65 & Kepler-62e & 1.61 & 1.19 & 4925 & 0.64 & \citet{Borucki2013}\\
701.04 & 18.1 & 1.42 & 0.41 & 4797 & 0.65 & Kepler-62f & 1.41 & 0.42 & 4925 & 0.64 & \citet{Borucki2013}\\
775.03 & 29.0 & 1.80 & 1.91 & 3898 & 0.54 & Kepler-52d & 1.95 & 2.81 & 4263 & 0.56 & \citet{Rowe2014}\\
812.03 & 28.1 & 1.94 & 1.16 & 3887 & 0.48 & Kepler-235e & 2.22 & 1.96 & 4255 & 0.55 & \citet{Rowe2014}\\
854.01 & 30.6 & 1.96 & 0.64 & 3593 & 0.47 & \nodata & \nodata & \nodata & \nodata & \nodata & \nodata\\
947.01 & 54.6 & 1.88 & 1.80 & 3750 & 0.46 & \nodata & \nodata & \nodata & \nodata & \nodata & \nodata\\
\tablenotemark{a}1126.02 & 13.8 & 1.80 & 0.21 & 5209 & 0.59 & \nodata & \nodata & \nodata & \nodata & \nodata & \nodata\\
1422.02 & 34.4 & 1.65 & 1.73 & 3517 & 0.37 & Kepler-296d & 2.09 & 2.90 & 3740 & 0.48 & \citet{Barclay2015}\\
1422.04 & 17.0 & 1.23 & 0.37 & 3517 & 0.37 & Kepler-296f & 1.80 & 0.62 & 3740 & 0.48 & \citet{Barclay2015}\\
1422.05 & 14.0 & 1.08 & 0.84 & 3517 & 0.37 & Kepler-296e & 1.53 & 1.41 & 3740 & 0.48 & \citet{Barclay2015}\\
1681.04 & 10.6 & 0.77 & 1.63 & 3669 & 0.35 & \nodata & \nodata & \nodata & \nodata & \nodata & \nodata\\
1989.01 & 32.0 & 1.84 & 1.83 & 5804 & 0.84 & \nodata & \nodata & \nodata & \nodata & \nodata & \nodata\\
2124.01 & 21.6 & 1.00 & 1.84 & 4029 & 0.55 & \nodata & \nodata & \nodata & \nodata & \nodata & \nodata\\
2184.02 & 9.2 & 1.89 & 1.63 & 4893 & 0.65 & \nodata & \nodata & \nodata & \nodata & \nodata & \nodata\\
2418.01 & 16.7 & 1.12 & 0.35 & 3724 & 0.41 & \nodata & \nodata & \nodata & \nodata & \nodata & \nodata\\
2529.02 & 12.8 & 1.90 & 1.28 & 4299 & 0.51 & Kepler-436b & 2.73 & 1.69 & 4651 & 0.70 & \citet{Torres2015}\\
2626.01 & 16.2 & 1.12 & 0.65 & 3482 & 0.35 & \nodata & \nodata & \nodata & \nodata & \nodata & \nodata\\
2650.01 & 14.1 & 1.25 & 1.14 & 3735 & 0.40 & Kepler-395c & 1.32 & 2.97 & 4262 & 0.56 & \citet{Rowe2014}\\
2719.02 & 14.0 & 1.72 & 1.99 & 4827 & 0.82 & \nodata & \nodata & \nodata & \nodata & \nodata & \nodata\\
3010.01 & 16.6 & 1.56 & 0.93 & 3903 & 0.52 & \nodata & \nodata & \nodata & \nodata & \nodata & \nodata\\
3138.01 & 10.8 & 0.57 & 0.47 & 2703 & 0.12 & \nodata & \nodata & \nodata & \nodata & \nodata & \nodata\\
3255.01 & 27.0 & 1.37 & 1.78 & 4427 & 0.62 & Kepler-437b & 2.14 & 2.15 & 4551 & 0.68 & \citet{Torres2015}\\
3282.01 & 17.9 & 1.97 & 1.30 & 3894 & 0.54 & \nodata & \nodata & \nodata & \nodata & \nodata & \nodata\\
3284.01 & 16.4 & 0.98 & 1.31 & 3688 & 0.46 & Kepler-438b & 1.12 & 1.40 & 3748 & 0.52 & \citet{Torres2015}\\
4036.01 & 25.6 & 1.83 & 1.02 & 4893 & 0.76 & \nodata & \nodata & \nodata & \nodata & \nodata & \nodata\\
4054.01 & 27.3 & 1.99 & 1.41 & 5380 & 0.78 & \nodata & \nodata & \nodata & \nodata & \nodata & \nodata\\
4060.01 & 27.3 & 1.96 & 1.82 & 5984 & 0.89 & \nodata & \nodata & \nodata & \nodata & \nodata & \nodata\\
4087.01 & 23.9 & 1.47 & 0.39 & 3813 & 0.48 & Kepler-440b & 1.86 & 1.20 & 4134 & 0.56 & \citet{Torres2015}\\
4356.01 & 16.5 & 1.91 & 0.29 & 4366 & 0.46 & \nodata & \nodata & \nodata & \nodata & \nodata & \nodata\\
4427.01 & 13.7 & 1.47 & 0.17 & 3668 & 0.43 & \nodata & \nodata & \nodata & \nodata & \nodata & \nodata\\
4450.01 & 15.1 & 1.98 & 1.38 & 5536 & 0.82 & \nodata & \nodata & \nodata & \nodata & \nodata & \nodata\\
4550.01 & 12.5 & 1.73 & 1.04 & 4771 & 0.70 & \nodata & \nodata & \nodata & \nodata & \nodata & \nodata\\
4622.01 & 13.9 & 1.93 & 0.34 & 4243 & 0.63 & Kepler-441b & 1.64 & 0.21 & 4340 & 0.55 & \citet{Torres2015}\\
4742.01 & 12.5 & 1.56 & 1.08 & 4569 & 0.65 & Kepler-442b & 1.34 & 0.66 & 4402 & 0.60 & \citet{Torres2015}\\
5202.01 & 8.8 & 1.83 & 0.63 & 6014 & 0.96 & \nodata & \nodata & \nodata & \nodata & \nodata & \nodata\\
5236.01 & 22.5 & 1.98 & 0.79 & 6241 & 1.03 & \nodata & \nodata & \nodata & \nodata & \nodata & \nodata\\
\tablenotemark{b}5475.01 & 28.0 & 1.66 & 0.68 & 6070 & 0.81 & \nodata & \nodata & \nodata & \nodata & \nodata & \nodata\\
5856.01 & 12.7 & 1.70 & 1.47 & 5906 & 0.85 & \nodata & \nodata & \nodata & \nodata & \nodata & \nodata\\
\tablenotemark{c}6343.01 & 9.9 & 1.90 & 0.61 & 6117 & 0.95 & \nodata & \nodata & \nodata & \nodata & \nodata & \nodata\\
\tablenotemark{c}6425.01 & 8.7 & 1.50 & 0.68 & 5942 & 0.95 & \nodata & \nodata & \nodata & \nodata & \nodata & \nodata\\
6676.01 & 10.3 & 1.81 & 1.18 & 6553 & 0.96 & \nodata & \nodata & \nodata & \nodata & \nodata & \nodata\\
6971.01 & 12.2 & 1.60 & 1.66 & 4989 & 0.79 & \nodata & \nodata & \nodata & \nodata & \nodata & \nodata\\
7016.01 & 11.8 & 1.13 & 0.56 & 5578 & 0.79 & Kepler-452b & 1.63 & 1.10 & 5757 & 1.11 & \citet{Jenkins2015}\\
7179.01 & 8.2 & 1.18 & 1.29 & 5845 & 1.20 & \nodata & \nodata & \nodata & \nodata & \nodata & \nodata\\
7223.01 & 9.1 & 1.53 & 0.57 & 5370 & 0.73 & \nodata & \nodata & \nodata & \nodata & \nodata & \nodata\\
\tablenotemark{c}7235.01 & 8.6 & 1.15 & 0.75 & 5606 & 0.76 & \nodata & \nodata & \nodata & \nodata & \nodata & \nodata\\
\tablenotemark{c}7470.01 & 8.9 & 1.90 & 0.60 & 5128 & 0.99 & \nodata & \nodata & \nodata & \nodata & \nodata & \nodata\\
\tablenotemark{c}7554.01 & 8.1 & 1.98 & 1.12 & 6315 & 1.09 & \nodata & \nodata & \nodata & \nodata & \nodata & \nodata\\
\tablenotemark{c}7567.01 & 11.3 & 1.46 & 0.10 & 4486 & 0.65 & \nodata & \nodata & \nodata & \nodata & \nodata & \nodata\\
\tablenotemark{c}7591.01 & 8.2 & 1.30 & 0.33 & 4906 & 0.67 & \nodata & \nodata & \nodata & \nodata & \nodata & \nodata\\
7592.01 & 10.4 & 1.55 & 0.07 & 3761 & 0.53 & \nodata & \nodata & \nodata & \nodata & \nodata & \nodata

\enddata
\tablecomments{KOIs with confirmed planet numbers have been confirmed as planetary in nature either via ground-based follow-up observations or statistical analyses. In these cases we list the confirmed \kepler{} planet number, confirmed values for the planet's radius, insolation flux, and stellar effective temperature and radius, and reference for the confirmation study.}
\tablenotetext{a}{Known to be a false positive via manual inspection.}
\tablenotetext{b}{Modeled at twice the true orbital period.}
\tablenotetext{c}{Likely to be a FP due to low-amplitude systematics given detailed manual vetting of the PDC light curves.}
\label{hzearthstab}
\end{deluxetable*}

There are a number of PCs in Table~\ref{hzearthstab} that are new in the Q1--Q17~DR24 catalog. A much larger fraction of them orbit solar-like stars compared to previously known PCs in the table, as well as having lower insolation flux values. Specifically, KOIs~6343.01, 6425.01, 7016.01, 7223.01, 7235.01, and 7470.01 have inferred radii between 1.13--1.90~\re{}, insolation fluxes between 0.56--0.75~\se{}, and orbit stars with \tstar{} between 5128--6117~K. We first note that they are generally also at lower SNR compared to previously known PCs, which coupled with being in single systems, puts them at higher risk for being undetected, low SNR, false positives. We also note that if any of them are confirmed to be planets by subsequent observations and analyses, their resulting radii and insolation fluxes could change significantly as a result of more accurate stellar parameters. However, the fact that there are a significant number of new PCs that orbit Sun-like stars and have insolation fluxes even less than that of Earth's represents great progress by the \kepler{} mission in determining the fraction of Earth-size planets in the habitable zone of Sun-like stars. In order to facilitate the prioritization of follow-up observations, we examine each PC from Table~\ref{hzearthstab} that is new in the Q1--Q17~DR24 catalog in detail in the subsections below. We utilize both the TCERT vetting forms \citep[][publicly available for every Q1--Q17~DR24 TCE at the Exoplanet Archive]{Coughlin2015a}, as well as the PDC data from MAST.

\subsubsection{KOI~1126.02}

KOI~1126.02 is a new KOI not correctly dispositioned by the robovetter. KOI 1126 (KIC 006307521) is contaminated by a nearby EB with a period of 29.745 days and a clearly visible secondary that is about half as deep as the primary. The first TCE produced by the \kepler{} pipeline, 006307521-01, which federates to KOI 1126.01, is detected at a period of 29.745 days, and the robovetter correctly dispositions it as a false positive due to a significant secondary, centroid offset, and ephemeris match. After removing the primary transits, the \kepler{} pipeline re-searched the data and detected a second TCE, 006307521-02, at a period of 475.954 days, or $\sim$16 times the first TCE's and EB's period, corresponding to a subset of just three of the EB's secondary eclipses. 

As it did appear transit-like, TCE 006307521-02 was designated as new KOI 1126.02. However, the detected period was off enough from an exact 16:1 ratio that it just barely failed to period match to either the previous TCE or the parent EB. Also, the robovetter centroid module has a safeguard to protect low SNR planet candidates, where no TCE is designated a false positive if it does not have at least three valid centroid measurements. The middle of the three events for 006307521-02 / KOI 1126.02 fell close enough to a data gap to prevent a valid centroid measurement, and thus the object was passed by the centroid module. However, we note that the Q1--Q17~DR24 KOI catalog prioritizes uniformity over the accuracy of individual targets, and this example shows why it is prudent to manually inspect the Q1--Q17~DR24 TCERT vetting forms \citep{Coughlin2015a}, before committing precious telescope time to observing individual high value targets.

\subsubsection{KOI~1681.04}

KOI~1681.04 appears to be a strong planet candidate with an inferred sub-Earth-size of 0.77 \re{} in a $\sim$22 day orbit around a late M-dwarf (0.35 \rsun, 3669K), resulting in an insolation flux 1.6 times that of Earth. There are three previously known planet candidates in this system, with shorter periods and radii of 0.69, 0.71, and 0.99~\re{}. The existence of this new candidate in a multi-planet system lends higher confidence to it being a real planet \citep{Rowe2014}. This new candidate has also been recently detected and published by \citet{Dressing2015}.

\subsubsection{KOI~2719.02}

KOI~2719.02 was first identified as a KOI in the Q1--Q16 KOI catalog \citep{Mullally2015a}, but was considered not transit-like and dispositioned as a FP. KOI~2719.02 was re-detected as a Q1--Q17~DR24 TCE and dispositioned as a PC by the robovetter. Manually examining the TCERT diagnostics, KOI~2719.02 does indeed appear to be a strong planet candidate. It is possible that detrending differences are responsible for the vetting differences between catalogs, as the star does have strong variability, though not near the periods of either KOI in the system. With a radius of 1.72~\re{} and an insolation flux of 1.99~\se{}, given its period of 106 days around a 0.82~\rsun{}, 4827~K star, it is likely KOI~2719.02 lies interior to the habitable zone, but still forms part of an interesting multi-planet system given that the inner candidate, KOI~2179.01, has a nearly identical size of 1.71~\re{}, though with an insolation flux of 152~\se{}.

\subsubsection{KOI~5475.01}

KOI~5475.01 was first detected as a Q1--Q16 TCE at a period of 448~days, and dispositioned as a FP due to a significant secondary \citep{Mullally2015a}. In the Q1--Q17~DR24 catalog, KOI~5475.01 was detected as a TCE at a period of 224 days and was dispositioned by the robovetter as a PC. Manual inspection confirms there is no discernible odd-even difference in Q1--Q17~DR24, and thus the Q1--Q16 TCE detection was at twice the true orbital period, resulting in a perceived secondary of identical depth and width at a phase of 0.5 during the Q1--Q16 vetting. While the period of this candidate is 224 days, it was first identified in Q1--Q16 and modeled with a period of 448~days (see \S\ref{koimodelsec}). Thus, the resulting insolation flux value of 0.68~\se{} given in the Q1--Q17~DR24 catalog is too low, and should actually be 1.71~\se{}, while the radius of 1.66~\re{} is still correct. This candidate also forms part of an interesting multi-planet system with the inner candidate, KOI~5475.02, at a radius of 0.54~\re{} and insolation flux of 230~\se{}, both around a 0.81~\rsun{}, 6070~K host star.

\subsubsection{KOI~6343.01}

KOI~6343.01 is a newly detected single PC in Q1--Q17~DR24, with an inferred size of 1.90~\re{} and insolation flux of 0.61~\se{}, given its period of 569 days around a 0.95~\rsun{}, 6117~K star. Manual inspection of its PDC light curve reveals that of its three transit events, the second event is likely a low amplitude SPSD, and the other two events may be smaller amplitude systematic events. The SPSD feature is not as readily visible in either the DV or alternate detrending, though it is perhaps not surprising given the KOI's low SNR of 9.9. We note the value of the Marshall metric is 8.1 for KOI~6343.01, which is very close to the threshold value of 10.0, above which the robovetter classifies TCEs/KOIs as FPs due to systematic events. For clarity, in Figure~\ref{marshallfig} we plot the distribution of the Marshall metric for the injTCEs (see \S\ref{injectsec}), where it can be seen that very few injected transiting planets have Marshall metrics near 10 or higher, even at low SNR. Overall, we deem it very likely that this object is actually a result of low amplitude systematics.

\subsubsection{KOI~6425.01}

KOI~6425.01 is a newly detected single PC in Q1--Q17~DR24, with an inferred size of 1.50~\re{} and insolation flux of 0.68~\se{}, given its period of 521 days around a 0.95~\rsun{}, 5942~K star. Manual inspection of its PDC light curve reveals that of its three transit events, the first event (in Q2) is likely due to a low amplitude SPSD, and the second event (in Q7) may be due to an edge effect. We note the value of the Marshall metric is 7.8 for KOI~6425.01, which is very close to the threshold value of 10.0, above which the robovetter classifies TCEs/KOIs as FPs due to systematic events.

\subsubsection{KOI~6676.01}

KOI~6676.01 is a newly detected single PC in Q1--Q17~DR24, with an inferred size of 1.81~\re{} and insolation flux of 1.18~\se{}, given its period of 439 days around a 0.96~\rsun{}, 6553~K star. Manual inspection of its PDC light curve reveals no systemic source of the signal for any of its three transits, and its Marshall metric value is 0.95, well below the FP threshold of 10.0. Thus KOI~6676.01 appears to be a reliable planet candidate, though we note it has a SNR of 10.3.

\begin{figure}
\centering
\includegraphics[width=\linewidth]{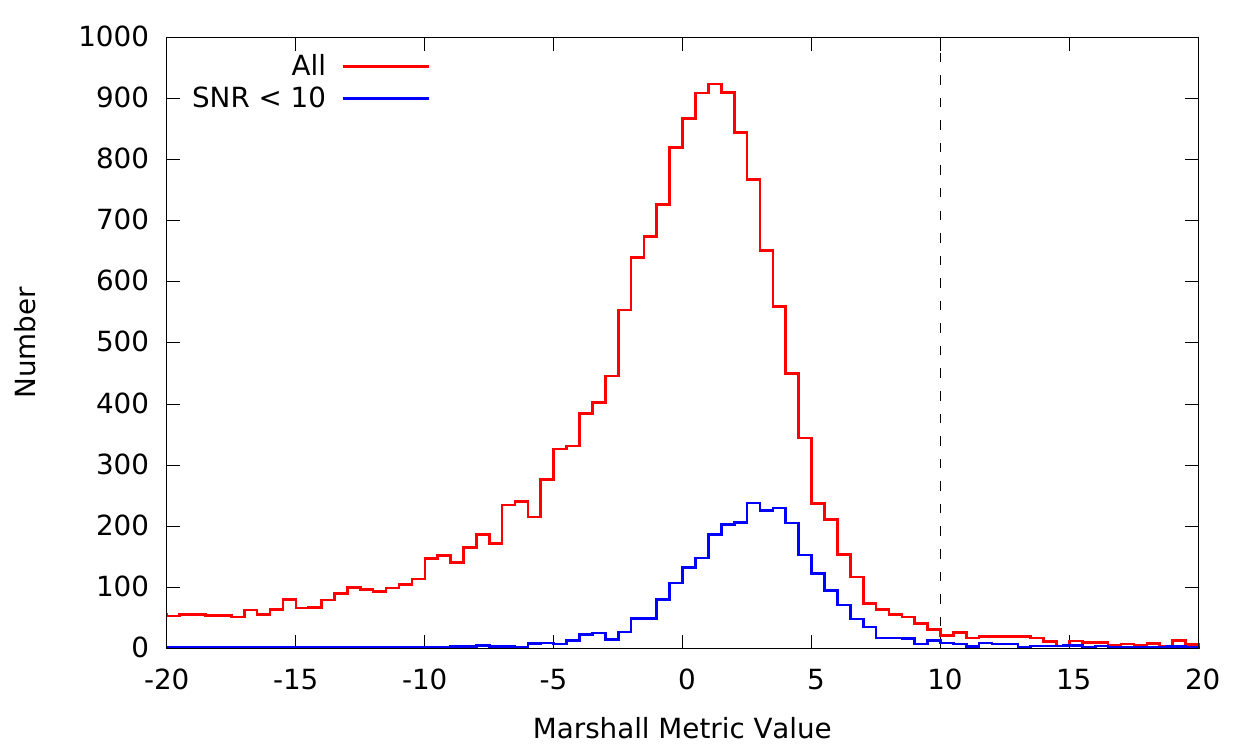}
\caption{The distribution of Marshall metric values for the injected TCEs. The red line represents all injected TCEs with computed Marshall metrics, while the blue line represents the subset of those with a SNR less than 10. The vertical dashed line represents the value above which the robovetter dispositions TCEs as FPs. Note that very few injected TCEs have Marshall values above this cutoff value.}
\label{marshallfig}
\end{figure}

\subsubsection{KOI~6971.01}

KOI~6971.01 is a newly detected single PC in Q1--Q17~DR24, with an inferred size of 1.60~\re{} and insolation flux of 1.66~\se{}, given its period of 129 days around a 0.79~\rsun{}, 4989~K star. Manual inspection reveals this to be a strong planet candidate, with 10 observed transits.

\subsubsection{KOI~7016.01}

KOI~7016.01 is a newly detected single PC in Q1--Q17~DR24, with an inferred size of 1.19~\re{} and insolation flux of 0.56~\se{}, given its period of 385 days around a 0.79~\rsun{}, 5578~K star. Given these catalog parameters it represents one of the most Earth-like planet candidates in the sample, at least in terms of size, insolation flux, and solar type host star. This KOI was recently designated Kepler-452b as it was validated by \citet{Jenkins2015} via spectroscopic follow-up of the host star and statistical analyses. However, they found that due to the host star being a more evolved star than previously indicated, the planet actually has a radius of 1.6~\re{} and an insolation flux of 1.1~\se{}.

\subsubsection{KOI~7179.01}

KOI~7179.01 is a newly detected single PC in Q1--Q17~DR24, with an inferred size of 1.18~\re{} and insolation flux of 1.29~\se{}, given its period of 407 days around a 1.2~\rsun{}, 5845~K star. Manual inspection reveals no evidence for any of its three transit events being due to systematics, though the KOI has a very low SNR of 8.2, so it is difficult to definitely discern the shape of individual events. KOI~7179.01 has a Marshall metric value of 5.9, which is a moderate value only 1.2$\sigma$ from the peak of the low SNR injTCE distribution (see Figure~\ref{marshallfig}). Overall this appears to be a good, but low SNR, PC with Earth-like size, insolation flux, and solar type host star.

\subsubsection{KOI~7223.01}

KOI~7223.01 is a newly detected single PC in Q1--Q17~DR24, with an inferred size of 1.53~\re{} and insolation flux of 0.57~\se{}, given its period of 317 days around a 0.73~\rsun{}, 5370~K star. Manual inspection reveals this to be a strong planet candidate, with 5 observed transits. KOI~7223.01 represents another new, possibly rocky planet candidate in the habitable zone of a late G-type star.

\subsubsection{KOI~7235.01}

KOI~7235.01 is a newly detected single PC in Q1--Q17~DR24, with an inferred size of 1.15~\re{} and insolation flux of 0.75~\se{}, given its period of 300 days around a 0.76~\rsun{}, 5606~K star. Manual inspection of its PDC light curve reveals that of its five transit events, two of them occur on the edges of gaps. Of the remaining three, one or two are maybe due to a low amplitude SPSD, though it is difficult to be sure given the KOI's low SNR of 9.1. KOI~7235.01 has a fairly high value for the Marshall metric of 8.1, close to the 10.0 FP threshold. Overall, we deem it most likely that this object is due to low amplitude systematics.

\subsubsection{KOI~7470.01}

KOI~7470.01 is a newly detected single PC in Q1--Q17~DR24, with an inferred size of 1.90~\re{} and insolation flux of 0.60~\se{}, given its period of 393 days around a 0.99~\rsun{}, 5128~K star. Manual inspection of its PDC light curve reveals that of its three transit events, the middle event (in Q9) is very likely due to a SPSD or step-wise discontinuity. Also, the value of the Marshall metric is 8.2, close to the 10.0 FP threshold. Overall, we deem it very likely that this object is actually a result of low amplitude systematics.

\subsubsection{KOI~7554.01}

KOI~7554.01 is a newly detected single PC in Q1--Q17~DR24, with an inferred size of 1.98~\re{} and insolation flux of 1.12~\se{}, given its period of 483 days around a 1.09~\rsun{}, 6315~K star. Manual inspection of its PDC light curve reveals that of its three transit events, the last event (in Q14) is very likely due to a SPSD. The value of the Marshall metric for KOI~7554.01 is 5.1. Overall, we deem it likely that this object is actually a result of low amplitude systematics.

\subsubsection{KOI~7567.01}

KOI~7567.01 is a newly detected single PC in Q1--Q17~DR24, with an inferred size of 1.46~\re{} and insolation flux of 0.10~\se{}, given its period of 608 days around a 0.65~\rsun{}, 4486~K star. Manual inspection of its PDC light curve reveals that of its three transit events, the first event (in Q1) is very likely due to a SPSD. The value of the Marshall metric is 9.5, very close to the 10.0 FP threshold. Overall, we deem it very likely that this object is actually a result of low amplitude systematics.

\subsubsection{KOI~7591.01}

KOI~7591.01 is a newly detected single PC in Q1--Q17~DR24, with an inferred size of 1.30~\re{} and insolation flux of 0.33~\se{}, given its period of 328 days around a 0.67~\rsun{}, 4906~K star. Manual inspection of its PDC light curve reveals that of its three transit events, the second event (in Q5) is likely due to a SPSD. The value of the Marshall metric is 6.7, which is moderately high. Overall, we deem it likely that this object is actually a result of low amplitude systematics.

\subsubsection{KOI~7592.01}

KOI~7592.01 is a newly detected single PC in Q1--Q17~DR24, with an inferred size of 1.55~\re{} and insolation flux of 0.07~\se{}, given its period of 382 days around a 0.53~\rsun{}, 3761~K star. Manual inspection reveals no evidence for any of its three transit events being due to systematics, though this KOI has a low SNR of 10.4, so it is difficult to definitively discern the shape of its 3 transit events. KOI~7592.01 has a Marshall metric value of 8.4, which is fairly high given the FP threshold of 10.0, but still only 2.2$\sigma$ from the median Marshall value for low SNR injTCEs (see Figure~\ref{marshallfig}). Overall this appears to be a borderline, low SNR planet candidate. If the signal really is due to a planet, it would be quite unique as it is the candidate with the lowest insolation flux in the Q1--Q17~DR24 catalog, given its long period and M-dwarf host star.

\section{Discussion}
\label{discusssec}

The Q1--Q17~DR24 KOI catalog represents the first time that every TCE from a \kepler{} pipeline search has been uniformly vetted. Of the \ntces{} Q1--Q17~DR24 TCEs, the robovetter ruled \nntltces{} as not transit-like, and another \nfpkoitces{} as transit-like false positives, leaving \npctces{} PCs. (Note that 5 of the TCEs that were designated as PCs were not given KOI numbers, as discussed in \S\ref{koisec}, resulting in a total of \npckois{} PCs in the Q1--Q17~DR24 catalog.) Combining these results with previous \kepler{} catalogs, there are now \ntotpcs{} PCs in the cumulative \kepler{} KOI catalog. Due to the uniform vetting, the vast majority of known KOIs from previous catalogs have been re-vetted, many of which were previously vetted with only a few quarters of \kepler{} data. This should result in more accurate dispositions for most KOIs, though we note that the catalog values uniformity over individual correctness. Users of the catalog who are interested in investigating individual KOIs are encouraged to check the disposition from this catalog, as well as the dispositions given by previous catalogs.

As only known contact eclipsing binaries were excluded from the \kepler pipeline transit search, and as the robovetter designates specific categories of false positives, the catalog is also a valuable repository of information for detached eclipsing binaries and other specific classes of false positives. For example, there are 1,215 on-target eclipsing binaries in the Q1--Q17~DR24 KOI catalog, which can be identified as those KOIs that were dispositioned FP only due to a significant secondary (i.e., no centroid offset nor ephemeris match was identified). The study of these EBs can yield valuable stellar science, especially when coupled with follow-up observations. There are 1,730 KOIs dispositioned FP due to a centroid offset or ephemeris match, which is a valuable sample for studying how \kepler{} targets are contaminated across the entire field. A third category of interest are KOIs that had visible secondary eclipses that could be attributed to planetary reflection and/or thermal emission, identified as PCs with the significant secondary flag marked, of which 40 exist in the catalog.

\begin{figure*}
\centering
\begin{tabular}{c}
\includegraphics[width=\linewidth]{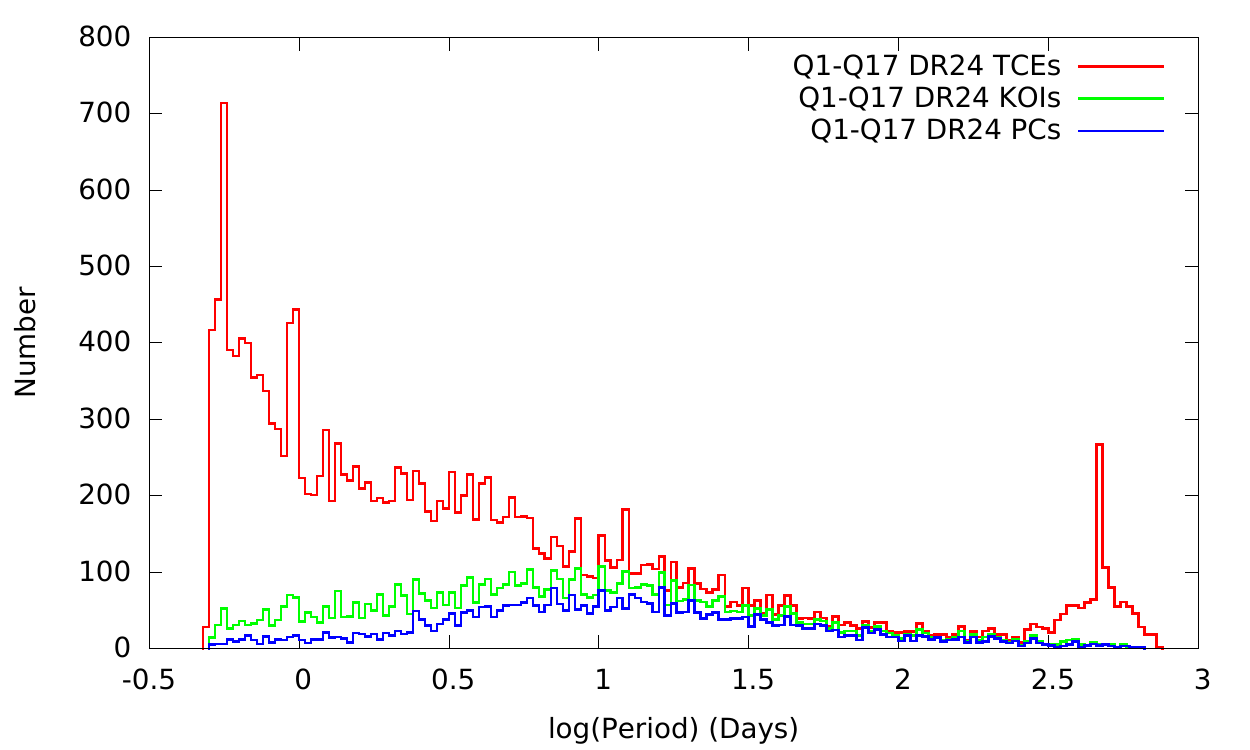}\\
\includegraphics[width=\linewidth]{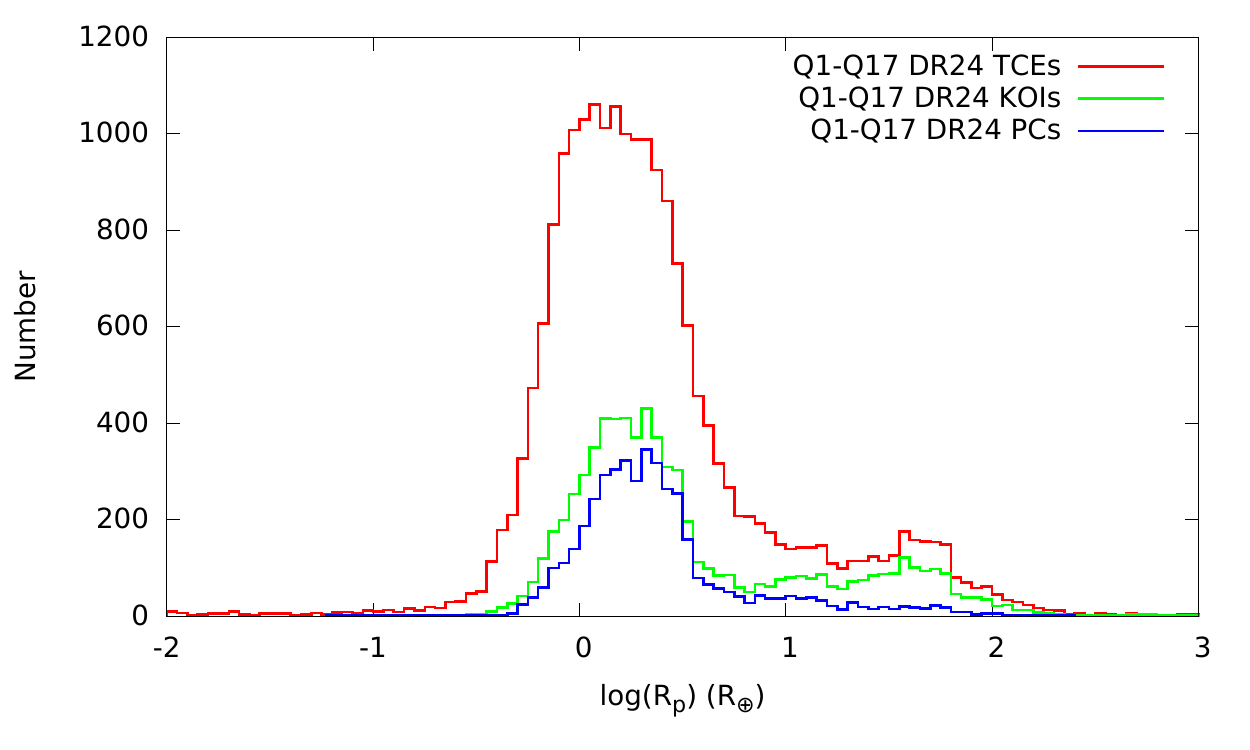}
\end{tabular}
\caption{The distribution of Q1--Q17~DR24 TCEs (red), KOIs (green), and PCs (blue) as a function of period (top) and radius (bottom). Note that the difference between the red and green lines represents the population of not transit-like FPs, and the difference between the green and blue lines represents the population of the transit-like FPs. Also note that, as shown, some planet candidates can have very large inferred radii, as FPs are purposely not designated based on depth or inferred size alone.}
\label{perhistfig}
\end{figure*}

In Figure~\ref{perhistfig} we plot a histogram of the number of Q1--Q17~DR24 TCEs, the number of TCEs designated as transit-like (KOIs), and the number of KOIs designated as PC as a function of period (similar to Figure~\ref{tce-comp-fig}) and planetary radius. As can be seen, the short- and long-period TCE excesses, as well as the local TCE period spikes, have been generally eliminated. The TCEs with very small and very large radii are also eliminated. FP KOIs, which are represented by the difference between the green and blue lines, must either have a significant secondary, centroid offset, or ephemeris match, and thus are principally due to eclipsing binaries. As expected, the FP KOI population is dominated by short periods and large radii, similar to the \kepler{} eclipsing binary period and radius distribution \citep{Prsa2011,Slawson2011}.

This is also the first time that artificial transit injection has been used in both the development and evaluation of the \kepler{} pipeline and the TCERT vetting process. We note again that special care should be taken in computing occurrence rates using this catalog due to the period-dependent search performed by the Q1--Q17~DR24 \kepler{} pipeline as a result of the bootstrap veto (\S\ref{falsenegsec}). \kepler{} mission completeness and reliability products\footnote{http://exoplanetarchive.ipac.caltech.edu/docs/Kepler\_completeness\_reliability.html} should also be used in conjunction with the Q1--Q17~DR24 catalog when computing occurrence rates. Overall though, the uniform vetting via our robotic approach, coupled with the transit injection results, enables a more accurate computation of the number of Earth-size planets in the habitable zone of Sun-like stars. 

Overall, the robovetter is successful in robustly identifying false positives while retaining valid planet candidates. As presented in \S\ref{injectsec}, the robovetter has a small false negative rate as measured by artificial transit injection. Given the qualitatively efficient elimination of FP TCEs shown in Figure~\ref{perhistfig}, and the quantitative comparison to ancillary catalogs in \S\ref{ancilsec}, the robovetter also likely has an overall small false positive rate. We note again here that the robovetter purposely does not designate false positives based on transit depth or inferred planet size, in order to produce a uniform catalog that is agnostic of the stellar parameters --- planet candidates with inferred radii several times that of Jupiter and larger are very likely to be due to eclipsing binaries, and users of this catalog are encouraged to make cuts on the inferred radii where it is appropriate for their scientific objectives. We also note that the false positive rate is likely enhanced for very low SNR candidates ($\lesssim$10), as shown by the manual inspection of new, low SNR candidates in \S\ref{hzsec}. We stress that full simulations of false positives, alongside the existing simulated planet transits, are needed to fully quantify the false positive rate as a function of SNR and other parameters, and thus calculate accurate occurrence rates.

The robovetter has known areas where it could be improved. For example, there are a handful of slightly eccentric eclipsing binaries with nearly equal primary and secondary depths that are detected as TCEs at half the true orbital period. These are not detected by the current odd-even depth test, and thus a test to search for an odd-even epoch offset is needed. A small number of TCEs due to flux contamination are sometimes detected at an integer ratio of the true orbital period, and when seasonal depth variations are present, they can sometimes escape identification by the robovetter. In addition, planets with strong TTVs can be erroneously labeled as false positives, though the number of these systems is extremely low both due to their intrinsic occurrence rate as well as non-detection by the \kepler{} pipeline, which is not designed to detect non-periodic transits resulting from strong TTVs. Finally, there are a number of variable stars which generate TCEs at the same period as their variability. While the stellar variability is obvious in the PDC data, both detrendings can sometimes make the resulting TCE appear to be transit-like, and thus a test to detect these cases would be valuable. These issues will be examined, and the robovetter further improved, for the next \kepler{} planet candidate catalog.

\newpage

\section{Conclusion}

We produced, for the first time, a uniform planet candidate catalog based on the entire 48 month \kepler{} dataset. We developed a robotic vetting program that mimics the human decision-making process employed by previous catalogs to examine the periodic signals identified by the \kepler{} pipeline. Our robotic vetting approach is able to eliminate the vast majority of false positive signals, while simultaneously retaining greater than 98\% of artificially injected planets similar to Earth in size and insolation flux, and over 99\% of confirmed planets. Coupled with the injection of artificial transits, these advancements allow for a more accurate computation of the fraction of Earth-size planets in the habitable zone of Sun-like stars. We note that this robotic vetting approach can be readily applied to other large-scale photometric survey missions, such as K2 \citep{Howell2014}, TESS \citep{Ricker2015}, and LSST \citep{Ivezic2008}.

\acknowledgments

We thank the anonymous referee for a careful reading of the paper and comments which greatly helped to improve the readability of the paper. B.~Q. gratefully acknowledges support by an appointment to the NASA Postdoctoral Program at the Ames Research Center, administered by Oak Ridge Associated Universities through a contract with NASA. D.~H. acknowledges support by the Australian Research Council's Discovery Projects funding scheme (project number DE140101364) and support by the National Aeronautics and Space Administration under Grant NNX14AB92G issued through the \kepler{} Participating Scientist Program. This paper includes data collected by the \kepler{} mission. Funding for the \kepler{} mission is provided by the NASA Science Mission directorate. The authors acknowledge the efforts of the \kepler{} Mission team for obtaining the calibrated pixel, light curve, and data validation reports used in this publication, which were generated by the \kepler{} Mission science pipeline through the efforts of the \kepler{} Science Operations Center and Science Office. The \kepler{} Mission is lead by the project office at NASA Ames Research Center. Ball Aerospace built the \kepler{} photometer and spacecraft which is operated by the mission operations center at LASP. These data products are archived at the NASA Exoplanet Science Institute, which is operated by the California Institute of Technology, under contract with the National Aeronautics and Space Administration under the Exoplanet Exploration Program. This research has made use of NASA's Astrophysics Data System. Some of the data presented in this paper were obtained from the Mikulksi Archive for Space Telescopes (MAST). STScI is operated by the Association of Universities for Research in Astronomy, Inc., under NASA contract NAS5-26555. Support for MAST for non-HST data is provided by the NASA Office of Space Science via grant NNX09AF08G and by other grants and contracts.

\newpage

\appendix

\section{List of Acronyms}
\label{acroappendsec}

NASA missions like \kepler{} tend to accumulate a large number of acronyms. Hence, we provide a summary of those used in this paper for easy reference, along with their definitions.\\

\begin{itemize}
\item[] \textbf{BKJD}: Barycentric \kepler{} Julian Date: BKJD = BJD - 2454833.0
\item[] \textbf{DR}: Data Release
\item[] \textbf{DV}: Data Validation: The module of the \kepler{} pipeline that provides diagnostics for TCEs.
\item[] \textbf{EB}: Eclipsing Binary
\item[] \textbf{EBWG}: \keplers{} Eclipsing Binary Working Group
\item[] \textbf{FP}: False Positive
\item[] \textbf{FPWG}: \keplers{} False Positive Working Group
\item[] \textbf{HZ}: Habitable Zone: The region around a star where a planet could have surface temperatures that allow for the presence of liquid water.
\item[] \textbf{KIC}: \kepler{} Input Catalog: The catalog of stars in the \kepler{} field that was used for target selection.
\item[] \textbf{KOI}: \kepler{} Object of Interest: A unique identifier of a signal consistent with a transiting or eclipsing system.
\item[] \textbf{MCMC}: Markov Chain Monte Carlo
\item[] \textbf{MES}: Multiple Event Statistic: The signal-to-noise ratio for the detection of a TCE by the TPS module of the \kepler{} pipeline.
\item[] \textbf{PC}: Planet Candidate
\item[] \textbf{SES}: Single Event Statistic: The signal-to-noise ratio for the detection of an individual transit-like event by the TPS module of the \kepler{} pipeline.
\item[] \textbf{SNR}: Signal to Noise Ratio
\item[] \textbf{TCE}: Threshold Crossing Event: A series of periodic flux decrements consistent with the signal produced by a transiting planet.
\item[] \textbf{TCERT}: Threshold Crossing Event Review Team: A committee that reviews TCEs to identify false positives and planet candidates.
\item[] \textbf{TPS}: Transiting Planet Search: The module of the \kepler{} pipeline that searches for transits.
\item[] \textbf{TTV}: Transit Timing Variation: A deviation in the expected time of transit due to gravitational interaction in multi-planet systems.
\end{itemize}

\section{Robovetter Mnemonic Flags}
\label{minorflagsec}

In Table~\ref{robodispstab} we list mnemonic flags that describe the results of individual robovetter tests in the comments column. Here we describe the meaning of each mnemonic flag.

\begin{itemize}
\item[] \textbf{ALT\_ROBO\_ODD\_EVEN\_TEST\_FAIL}: The TCE failed the robovetter's odd-even depth test on the alternate detrending, and thus is marked as a FP due to a significant secondary.
\item[] \textbf{ALT\_SEC\_COULD\_BE\_DUE\_TO\_PLANET}: A significant secondary eclipse was detected in the alternate detrending, but it was determined to possibly be due to planetary reflection and/or thermal emission. While the significant secondary major flag remains set, the TCE is dispositioned as a PC.
\item[] \textbf{ALT\_SEC\_SAME\_DEPTH\_AS\_PRI\_COULD\_BE\_TWICE\_TRUE\_PERIOD}: A significant secondary eclipse was detected in the alternate detrending, but it was determined to be the same depth as the primary within the uncertainties. Thus, the TCE is possibly a PC that was detected at twice the true orbital period. When this flag is set, it acts as an override to other flags such that the significant secondary major flag is not set, and thus the TCE is dispositioned as a PC if no other major flags are set.
\item[] \textbf{ALT\_SIG\_PRI\_MINUS\_SIG\_POS\_TOO\_LOW}: The difference of the primary and positive event significances, computed by the model-shift test using the alternate detrending, is below the threshold $\sigma'_{\rm FA}$. This indicates the primary event is not unique in the phased light curve, and thus the TCE is dispositioned as a FP with the not transit-like major flag set.
\item[] \textbf{ALT\_SIG\_PRI\_MINUS\_SIG\_TER\_TOO\_LOW}: The difference of the primary and tertiary event significances, computed by the model-shift test using the alternate detrending, is below the threshold $\sigma'_{\rm FA}$. This indicates the primary event is not unique in the phased light curve, and thus the TCE is dispositioned as a FP with the not transit-like major flag set.
\item[] \textbf{ALT\_SIG\_PRI\_OVER\_FRED\_TOO\_LOW}: The significance of the primary event divided by the ratio of red noise to white noise in the light curve, computed by the model-shift test using the alternate detrending, is below the threshold $\sigma_{\rm FA}$. This indicates the primary event is not significant compared to the amount of systematic noise in the light curve, and thus the TCE is dispositioned as a FP with the not transit-like major flag set.
\item[] \textbf{CENTROID\_SIGNIF\_UNCERTAIN}: The significance of the centroid offset cannot be measured to high enough precision, and thus the centroid module can not confidently disposition the TCE as a FP. This is typically due to having only a very small number (3 or 4) of offset measurements, all with low SNR.
\item[] \textbf{CLEAR\_APO}: The TCE was marked as a FP due to a centroid offset because the transit occurs on a star that is spatially resolved from the target.
\item[] \textbf{CROWDED\_DIFF}: More than one potential stellar image was found in the difference image. The EYEBALL flag is always set when the CROWDED\_DIFF flag is set.
\item[] \textbf{DV\_ROBO\_ODD\_EVEN\_TEST\_FAIL}: The TCE failed the robovetter's odd-even depth test on the DV detrending, and thus is marked as a FP due to a significant secondary.
\item[] \textbf{DV\_SEC\_COULD\_BE\_DUE\_TO\_PLANET}: A significant secondary eclipse was detected in the DV detrending, but it was determined to possibly be due to planetary reflection and/or thermal emission. While the significant secondary major flag remains set, the TCE is dispositioned as a PC.  
\item[] \textbf{DV\_SEC\_SAME\_DEPTH\_AS\_PRI\_COULD\_BE\_TWICE\_TRUE\_PERIOD}: A significant secondary eclipse was detected in the DV detrending, but it was determined to be the same depth as the primary within the uncertainties. Thus, the TCE is possibly a PC that was detected at twice the true orbital period. When this flag is set, it acts as an override to other flags such that the significant secondary major flag is not set, and thus the TCE is dispositioned as a PC if no other major flags are set.
\item[] \textbf{DV\_SIG\_PRI\_MINUS\_SIG\_POS\_TOO\_LOW}: The difference of the primary and positive event significances, computed by the model-shift test using the DV detrending, is below the threshold $\sigma'_{\rm FA}$. This indicates the primary event is not unique in the phased light curve, and thus the TCE is dispositioned as a FP with the not transit-like major flag set.
\item[] \textbf{DV\_SIG\_PRI\_MINUS\_SIG\_TER\_TOO\_LOW}: The difference of the primary and tertiary event significances, computed by the model-shift test using the DV detrending, is below the threshold $\sigma'_{\rm FA}$. This indicates the primary event is not unique in the phased light curve, and thus the TCE is dispositioned as a FP with the not transit-like major flag set.  
\item[] \textbf{DV\_SIG\_PRI\_OVER\_FRED\_TOO\_LOW}: The significance of the primary event divided by the ratio of red noise to white noise in the light curve, computed by the model-shift test using the DV detrending, is below the threshold $\sigma_{\rm FA}$. This indicates the primary event is not significant compared to the amount of systematic noise in the light curve, and thus the TCE is dispositioned as a FP with the not transit-like major flag set.  
\item[] \textbf{EYEBALL}: The metrics used by the centroid module are very close to the decision boundaries, and thus the centroid disposition of this TCE is uncertain and warrants further scrutiny. No TCEs are marked as a FP due to a centroid offset if this flag is set.
\item[] \textbf{FIT\_FAILED}: The transit was not fit by a model in DV and thus no difference images were created for use by the centroid module. Thus, the TCE is not failed due to a centroid offset by default. This flag is typically set for very deep transits due to eclipsing binaries.
\item[] \textbf{INVERT\_DIFF}: One or more difference images were inverted, meaning the difference image claims the star got brighter during transit. This is usually due to variability of the target star and suggests the difference image should not be trusted. When this flag is set, the TCE is marked as a candidate that requires further scrutiny, i.e., the EYEBALL flag is set and the TCE is not marked as a FP due to a centroid offset.
\item[] \textbf{KIC\_OFFSET}: The centroid module measured the offset distance relative to the star's recorded position in the Kepler Input Catalog (KIC), not the out of transit centroid. The KIC position is less accurate in sparse fields, but more accurate in crowded fields. If this is the only flag set, there is no reason to believe a statistically significant centroid shift is present \citep{Mullally2015c}.
\item[] \textbf{LPP\_ALT\_TOO\_HIGH}: The LPP value \citep{Thompson2015b}, as computed using the alternate detrending, is above the robovetter threshold. This indicates the TCE is not transit-shaped, and thus is dispositioned as a FP with the not transit-like major flag set.
\item[] \textbf{LPP\_DV\_TOO\_HIGH}: The LPP value, as computed using the DV detrending, is above the robovetter threshold. This indicates the TCE is not transit-shaped, and thus is dispositioned as a FP with the not transit-like major flag set.  
\item[] \textbf{MARSHALL\_FAIL}: The TCE failed the Marshall metric \citep{Mullally2015b}, which indicates that the TCE's individual transits are not transit-shaped and more likely due to instrumental artifacts. Thus, the TCE is dispositioned as a FP with the not transit-like major flag set.
\item[] \textbf{OTHER\_TCE\_AT\_SAME\_PERIOD\_DIFF\_EPOCH}: Another TCE on the same target with a higher planet number was found to have the same period as the current TCE, but a significantly different epoch. This indicates the current TCE is an eclipsing binary with the other TCE representing the secondary eclipse. If the ALT\_SEC\_COULD\_BE\_DUE\_TO\_PLANET and DV\_SEC\_COULD\_BE\_DUE\_TO\_PLANET flags are not set, the TCE is dispositioned as a FP with the significant secondary major flag set.
\item[] \textbf{PARENT\_IS\_X}: The TCE has been identified as a FP due to an ephemeris match. This flag indicates the most likely parent, or true physical source of the signal, where X will be substituted for the parent's name. Note that X is not guaranteed to be the true parent, but simply is the most likely source given the information available.
\item[] \textbf{PERIOD\_ALIAS\_IN\_ALT\_DATA\_SEEN\_AT\_X:1}: Using the results of the model-shift test (specifically the phases of the primary, secondary, and tertiary events) a possible period alias is seen at X:1, where X is an integer. This indicates the TCE has likely been detected at a period that is X times longer than the true orbital period. This flag is currently informational only and not used to declare any TCE a FP.
\item[] \textbf{RESID\_OF\_PREV\_TCE}: The TCE has the same period and epoch as a previous transit-like TCE. This indicates the current TCE is simply a residual artifact of the previous TCE after it was removed from the light curve. Thus, the current TCE is dispositioned as a FP with the not transit-like major flag set.
\item[] \textbf{SAME\_P\_AS\_PREV\_NTL\_TCE}: The current TCE has the same period as a previous TCE that was dispositioned as FP with the not transit-like major flag set. This indicates that the current TCE is due to the same not transit-like signal. Thus, the current TCE is dispositioned as a FP with the not transit-like major flag set.
\item[] \textbf{SATURATED}: The star is saturated. The assumptions employed by the centroid robovetter module break down for saturated stars, so the TCE is marked as a candidate requiring further scrutiny, i.e., the EYEBALL flag is set and the TCE is not marked as a FP due to a centroid offset.
\item[] \textbf{SEASONAL\_DEPTH\_DIFFS\_IN\_ALT}: There appears to be a significant difference in the computed TCE depth when using the alternate detrending light curves from different seasons. This indicates significant light contamination is present, usually due to a bright star at the edge of the image, which may or may not be the source of the signal. As it is impossible to determine whether or not the TCE is on-target from this flag alone, it is currently informational only and not used to declare any TCE a FP.
\item[] \textbf{SEASONAL\_DEPTH\_DIFFS\_IN\_DV}: There appears to be a significant difference in the computed TCE depth when using the DV detrending light curves from different seasons. This indicates significant light contamination is present, usually due to a bright star at the edge of the image, which may or may not be the source of the signal. As it is impossible to determine whether or not the TCE is on-target from this flag alone, it is currently informational only and not used to declare any TCE a FP.  
\item[] \textbf{SIG\_SEC\_IN\_ALT\_MODEL\_SHIFT}: The significance of the secondary event divided by the ratio of red noise to white noise in the light curve, computed by the model-shift test using the alternate detrending, is above the threshold $\sigma_{\rm FA}$. Also, the difference between the secondary and tertiary event significances, and the difference between the secondary and positive event significances, both computed by the model-shift test using the alternate detrending, is above the threshold $\sigma'_{\rm FA}$. This indicates that there is a unique and significant secondary event in the light curve, i.e., a secondary eclipse. Thus, assuming the ALT\_SEC\_COULD\_BE\_DUE\_TO\_PLANET flag is not set, the TCE is dispositioned as a FP with the significant secondary flag set.
\item[] \textbf{SIG\_SEC\_IN\_DV\_MODEL\_SHIFT}: The significance of the secondary event divided by the ratio of red noise to white noise in the light curve, computed by the model-shift test using the DV detrending, is above the threshold $\sigma_{\rm FA}$. Also, the difference between the secondary and tertiary event significances, and the difference between the secondary and positive event significances, both computed by the model-shift test using the DV detrending, is above the threshold $\sigma'_{\rm FA}$. This indicates that there is a unique and significant secondary event in the light curve, i.e., a secondary eclipse. Thus, assuming the DV\_SEC\_COULD\_BE\_DUE\_TO\_PLANET flag is not set, the TCE is dispositioned as a FP with the significant secondary flag set.
\item[] \textbf{SIGNIF\_OFFSET}: There is a statistically significant shift in the centroid during transit. This indicates the variability is not due to the target star. Thus, the TCE is dispositioned as a FP with the centroid offset major flag set.
\item[] \textbf{THIS\_TCE\_IS\_A\_SEC}: The TCE is determined to have the same period, but different epoch, as a previous transit-like TCE. This indicates that the current TCE corresponds to the secondary eclipse of an eclipsing binary (or planet if the ALT\_SEC\_COULD\_BE\_DUE\_TO\_PLANET or DV\_SEC\_COULD\_BE\_DUE\_TO\_PLANET flags are set.) Thus, the current TCE is dispositioned as a FP with both the not transit-like and significant secondary major flags set.
\item[] \textbf{TOO\_FEW\_CENTROIDS}: The PRF centroid fit used by the centroid module does not always converge, even in high SNR difference images. This flag is set if centroid offsets are recorded for fewer than 3 high SNR difference images.
\item[] \textbf{TOO\_FEW\_QUARTERS}: Fewer than 3 difference images of sufficiently high SNR are available, and thus very few tests in the centroid module are applicable to the TCE. If this flag is set in conjunction with the CLEAR\_APO flag, the source of the transit may be on a star clearly resolved from the target.
\item[] \textbf{TRANSITS\_NOT\_CONSISTENT}: The TCE had a max\_ses\_in\_mes / mes ratio of greater than 0.9, and a period greater than 90 days. This indicates that the TCE is dominated by a single large event, and thus is due to a systematic feature such as a sudden pixel sensitivity dropout. Thus, the TCE is dispositioned as a FP with the not transit-like major flag set.
\end{itemize}


\bibliography{AstroRefs.bib}

\end{document}